\documentclass[preprint,12pt]{elsarticle}

\usepackage{amssymb}
\usepackage{amsmath}
\usepackage{array}
\usepackage{float}
\usepackage{url}
\journal{Computers in Biology and Medicine}

\begin{document}

\begin{frontmatter}



\title{Cross-Frequency Bispectral EEG Analysis of Reach-to-Grasp Planning and Execution}


% \author{Sima Ghafoori, Anna Cetera, Ali Rabiee, Mohammad Hasan Farhadi, Mariusz Furmanek, Yalda Shahriari, Reza Abiri} %% Author name
\author[label1]{Sima Ghafoori}
\author[label1]{Anna Cetera}
\author[label1]{Ali Rabiee}
\author[label1]{MH Farhadi}
\author[label1]{Rahul Singh}
\author[label2]{Mariusz Furmanek}
\author[label1]{Yalda Shahriari}
\author[label1]{Reza Abiri}

% \affiliation{organization={University of Rhode Island },%Department and Organization
%             % addressline={}, 
%             city={Kingston},
%             % postcode={}, 
%             state={RI},
%             country={USA}}

\affiliation[label1]{
  organization={Department of Electrical, Computer and Biomedical Engineering, University of Rhode Island},
  city={Kingston},
  state={RI},
  country={USA}
}

\affiliation[label2]{
  organization={Department of Physical Therapy, University of Rhode Island},
  city={Kingston},
  state={RI},
  country={USA}
}

%% Abstract
\begin{abstract}
Neural control of grasping emerges from nonlinear interactions across multiple
brain rhythms, yet EEG-based motor decoding has largely relied on linear,
second-order spectral measures. Here, we investigate whether higher-order
cross-frequency dynamics encode meaningful distinctions between motor planning
and execution execution during natural reach-to-grasp behavior. Using a
cue-based experimental paradigm, EEG was recorded during executed precision
and power grips, enabling stage-resolved analysis of preparatory and
execution-related neural activity.

Cross-frequency bispectral analysis was applied to compute complex bicoherence
matrices across canonical frequency band pairs, from which magnitude- and
phase-based features were extracted. Classification, permutation-based
feature selection, and within-subject statistical testing revealed that execution is associated with markedly stronger and more discriminative
nonlinear coupling than planning, with dominant contributions from $\beta$-
and $\gamma$-driven interactions. In contrast, decoding of precision versus
power grips showed comparable performance across stages (planning vs. execution), indicating that grasp-type representations emerge during planning and persist into execution. Exploratory single-feature analyses further identified focal, stage-dependent
modulation of nonlinear coupling in central motor regions.

Spectral and spatial analyses demonstrated that informative bispectral
features reflect coordinated activity across prefrontal, central, and
occipital areas, while substantial feature redundancy enables effective
dimensionality reduction without loss of performance. Together, these results
extend bispectral EEG analysis to ecologically valid grasp execution and highlight nonlinear cross-frequency coupling as a powerful and interpretable marker of motor planning and execution, with direct relevance to brain--computer interfaces and neuroprosthetic control.
\end{abstract}


% \begin{highlights}
% \end{highlights}

%% Keywords

\end{frontmatter}



\section{Introduction}
\label{sec1}
%% Labels are used to cross-reference an item using \ref command.
Electroencephalogram (EEG) is inherently nonlinear and non-Gaussian, yet traditional analytic methods predominantly rely on second-order measures such as power spectral density and coherence, which are limited to capturing only pairwise linear dependencies \cite{sigl1994introduction, 221324}. To uncover the richer dynamics present in neural activity, higher-order spectral techniques—collectively known as polyspectra—have been developed to characterize interactions across multiple frequency components simultaneously \cite{221324}. Among these, the bispectrum, or third-order spectrum, is particularly notable \cite{221324}. It preserves both magnitude and phase information and quantifies quadratic phase coupling between frequency components, offering insights into cross-frequency interactions and nonlinear processes that conventional methods typically miss \cite{sigl1994introduction, kovach2018bispectrum}. Although the bispectrum has its origins in statistical signal processing, its application in neuroscience is relatively recent, emerging as a powerful yet underutilized tool for probing the complex structure of brain activity \cite{bartz2024bispectral, chella2017non}. 

The utility of bispectral analysis has been demonstrated across a broad spectrum of neuroscientific and clinical contexts. In neurodegenerative research, bispectrum-derived features have proven effective in characterizing disease progression in Alzheimer’s disease, where altered cross-frequency interactions reliably differentiate patients from healthy controls \cite{maturana2020inter}. In clinical monitoring, bispectral indices extracted from EEG signals are widely employed to assess the depth of anesthesia, offering a robust, noninvasive metric for intraoperative evaluation \cite{myles2004bispectral, gan1997bispectral, 11077565, 10637197}. Bispectral methods have also been applied to epilepsy and Parkinson’s disease, where higher-order spectral features reveal pathological nonlinearities that elude conventional analyses \cite{sigl1994introduction, hyafil2015misidentifications}. Collectively, these diverse applications highlight the bispectrum’s capacity to uncover subtle neural dynamics with significant translational value, supporting its extension into emerging domains such as motor control and brain–computer interfacing \cite{kovach2018bispectrum}.

Beyond its clinical and diagnostic applications, the bispectrum has also gained attention in motor neuroscience, particularly within motor imagery (MI) studies for brain–computer interfacing (BCI) \cite{sun2019advanced, saikia2011bispectrum, kotoky2014bispectrum, shahid2011bispectrum, 11029483, 10782163, 10930540}. These studies have shown that higher-order spectral features can effectively discriminate imagined motor tasks—most commonly \cite{kotoky2014bispectrum}, simple left- versus right-hand executions. \cite{salwa2018classification}. Early research employed relatively basic metrics, such as the sum of log-bispectrum or simple statistical descriptors of bispectral magnitude, yielding moderate classification performance. Building on these foundations, later work introduced more sophisticated approaches, including a wider array of bispectral features, advanced bispectrum-based metrics, and data-driven channel selection strategies \cite{jin2020bispectrum}, collectively achieving classification accuracies of up to 90\% on public MI datasets \cite{das2016multiple}. More recent developments have integrated bispectral representations with deep learning architectures—such as hybrid convolutional neural network (CNN) and gated recurrent unit (GRU) models—to automatically learn nonlinear temporal–spectral patterns, reflecting the increasing methodological sophistication in this area \cite{liu2022bispectrum}. Related invasive studies using electrocorticography (ECoG) have further enhanced MI classification by combining bispectral features with execution-related potentials \cite{paul2017higher}. 

Despite recent advances in motor related studies, most of these studies have applied bispectral analysis within a relatively narrow feature–classification framework, relying primarily on statistical or entropy-based summaries of bispectral magnitude rather than conducting comprehensive cross-frequency analyses \cite{sun2019advanced, saikia2011bispectrum, kotoky2014bispectrum, shahid2011bispectrum, das2016multiple, jin2020bispectrum}. Consequently, the literature has largely focused on coarse motor distinctions, such as left- versus right-hand imagery, with limited investigation into more naturalistic or functionally relevant motor tasks. To the best of our knowledge, no prior studies have employed bispectral analysis to examine comprehensive movement tasks, such as executed reach-to-grasp in human EEG—such as differentiating between precision and power grips. Instead, applications of higher-order spectral methods in executed execution research have typically remained confined to broad motor class comparisons (e.g., left vs. right hand), leaving a critical gap in our understanding of how nonlinear cross-frequency dynamics contribute to the planning and execution of ecologically valid grasping behaviors. 

Building on this body of work, reach-to-grasp studies have employed a variety of recording modalities and analytic techniques, each offering complementary insights but also presenting notable limitations \cite{farhadi2025human}. Neuroimaging methods such as functional magnetic resonance imaging (fMRI) provide precise spatial localization of grasp-related cortical areas, yet their poor temporal resolution limits the ability to capture rapid neural dynamics. Invasive approaches, including ECoG and single-unit recordings, yield high-resolution signals but are restricted to clinical or animal models. In contrast, EEG offers a compelling balance—noninvasive, portable, and cost-effective—while enabling real-time monitoring of execution-related brain activity, making it particularly suited for translational applications in neuroprosthetics and rehabilitation \cite{bodda2022exploring}. Analytically, most EEG-based grasping studies have emphasized temporal features such as execution-related cortical potentials, or linear spectral measures like power spectral density and wavelet transforms \cite{xu2021decoding,schwarz2017decoding,schwarz2020analyzing,sburlea2021disentangling}. While these methods have advanced decoding accuracy, they are inherently limited in their ability to capture nonlinear neural dynamics. The potential of higher-order spectral techniques—such as bispectrum—to reveal cross-frequency interactions during executed grasping executions remains largely unexplored. 

Motor planning and execution are supported by distinct neural states that differ in both spectral content and cross-frequency coordination. Previous electrophysiological studies have shown that low-frequency rhythms (e.g., $\delta$ and $\theta$) are associated with large-scale integrative and anticipatory processes, whereas higher-frequency activity in the $\beta$ and $\gamma$ ranges is more closely linked to sensorimotor processing, execution, and local cortical computation \cite{pfurtscheller1999event,engel2010beta}. While many investigations have characterized these processes using band-limited power or linear coupling measures, growing evidence suggests that behaviorally relevant information is encoded in nonlinear interactions between frequency bands rather than within isolated rhythms \cite{canolty2010functional,jensen2012oscillatory}. Such cross-frequency interactions are inherently non-Gaussian and phase dependent, making them poorly captured by second-order statistics. Bispectral features, which are sensitive to quadratic phase coupling and waveform asymmetry, therefore provide candidate neurophysiological markers of task-dependent neural coordination that may distinguish planning and execution states beyond conventional spectral analyses \cite{kovach2018bispectrum,bartz2020beyond}.

This study leverages state-of-the-art bispectral analysis to investigate the neural mechanisms associated with two of the most common everyday grasp types: precision and power grip. To achieve this, we developed a custom platform and designed a cue-based experimental paradigm that ensures randomized, unbiased trials for accurate data collection. By extracting features from band-pair biocoherence matrices and leveraging both magnitude- and phase-based descriptors, our framework captures nonlinear cross-frequency interactions that conventional spectral and linear coupling methods overlook, providing candidate neurophysiological markers of task-related brain states. This work makes several key contributions: first, it extends bispectral analysis from the domain of motor imagery to the ecologically valid setting of executed reach-to-grasp; second, it demonstrates the discriminative potential of bispectrum-derived features for decoding natural grasp types; third, it advances understanding of the \textit{planning stage of reaching}, a critical but relatively understudied aspect of grasping; and fourth, it highlights the broader potential of bispectrum by opening the door to more advanced analyses, such as cross-bispectrum across channels for investigating network-level interactions. Collectively, these contributions establish a methodological and conceptual foundation for integrating higher-order spectral measures into neuroscience research, brain–computer interface development, and neuroprosthetic control. 

\section{Methods}
\label{sec2}

\subsection{Experimental Paradigm}
\begin{figure*}[!t]
\centering
\includegraphics[width=0.95\textwidth]{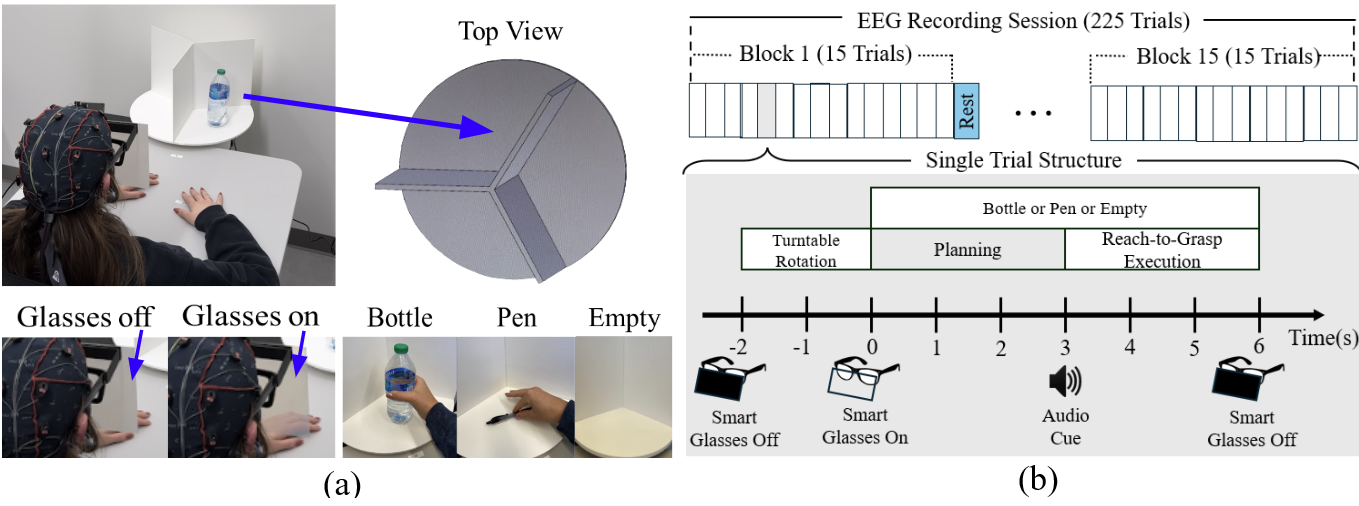}
\caption{Experimental Setup and Data Collection Protocol for Vision-Based Grasping Platform. (a) Depicts a participant wearing the EEG recording system during the experimental protocol. The top view of the turntable rotates to present one of three conditions at random: bottle, pen, or empty. (b) Single trial structure and outline of EEG recording session, beginning with the planning stage then the reach-to-grasp execution following audio cue onset. }
\label{setup}
\end{figure*}

Our experimental setup features a custom-designed, motorized 3D turntable divided into three sections to present two objects—a bottle and a pen—and a no-object scenario. The turntable was controlled via a computer using a TB6600 4A 9-42V Stepper Motor Driver paired with a Nema 17 Stepper Motor (1.7A Bipolar). Custom Python software was developed to ensure seamless synchronization of hardware and software components, as well as to log events critical for accurate EEG signal processing. Participants wore PC-controlled ``smart eyeglasses'' equipped with a smart film that alternated between transparent and opaque states. This feature was essential for managing object visibility and eliminating potential bias in participants’ reactions. During the experiment, participants were seated comfortably in a neutral posture, with their palms resting down approximately 30 cm from the center of the object display. As illustrated in Figure \ref{setup}-b, participants observed the setup for three seconds (planning stage), followed by an auditory cue (a buzzing sound) signaling the start of the grasping stage, which was limited to three seconds (execution stage). After each grasping stage, the turntable rotated to present the next condition (either an object or the no-object scenario). During these transitions, the smart eyeglasses turned opaque to prevent anticipatory bias, and its latency was measured to be < 10 ms, resulting in a delay that is negligible relative to the planning window. Moreover, visual inspection of the Fp1 and Fp2 EEG channels confirmed that no observable electrical artifacts were induced by the voltage switching of the smart glasses. Figure \ref{setup} provides a detailed depiction of the experimental platform and paradigm \cite{cetera2025macroscopic}.

\subsection{EEG Data Acquisition and Preprocessing}
EEG recordings were conducted using a biosignal amplifier equipped with 16 active electrodes (g.tec medical engineering, Austria). The electrodes were strategically placed across all brain regions according to the 10-20 international system and grouped by functional region: prefrontal attentional sites FP1 and FP2; frontal executive/motor-planning sites F3, Fz and F4; the central motor strip comprising primary-motor electrodes C3, Cz and C4 together with temporal–somatosensory sites T7 and T8; parietal sensor-integration electrodes P3, Pz and P4; and parieto-occipital visual-feedback sites PO7, PO8 and Oz. This montage provides continuous coverage from high-level decision and planning areas through the primary sensorimotor cortex to posterior visual regions. 

EEG data were recorded at a sampling rate of 256 Hz and subsequently filtered offline using a zero-phase, 4th-order Butterworth bandpass filter with cutoff frequencies set between 0.5 and 40 Hz. To minimize eye execution artifacts, participants maintained a fixed gaze along a consistent visual axis toward the presented object during each trial. Eye blinks and muscle artifacts were removed using FastICA (Fast Independent Component Analysis), with components identified and rejected through visual inspection based on characteristic artifact patterns. Following preprocessing, single-trial EEG data were sorted by class labels: pen (class ID: 0), bottle (class ID: 1), and empty (class ID: 2). For all subsequent feature extraction procedures, zero-phase, 4th-order Butterworth filters with cutoff frequencies (lower: 0.1 Hz, higher: 40 Hz) were consistently employed for both bandpass and low-pass filtering.

\subsection{Data Collection Protocol}

Ten healthy participants (5 males and 5 females) aged 21 to 40 were recruited from the University of Rhode Island (URI) for this study. All participants reported no history of motor deficits or neurological impairments. Before the experiment, they were fully informed about the protocol, signed an informed consent form, and received monetary compensation for their participation. The study was approved by the Institutional Review Board (IRB) of the University of Rhode Island (IRB \#1944644). 

For each participant, 75 trials were collected per condition (empty, pen, and bottle). Each run comprised three randomized trials, with the number of runs per block adjusted based on the participant's comfort. To minimize fatigue and maintain attention, no block exceeded 15 runs, and participants were offered short breaks between blocks as needed.

\subsection{Bispectral Analysis}

The bispectrum is a higher-order frequency-domain analysis tool that extends the concept of the power spectrum to capture phase coupling and non-linear interactions within a signal \cite{nikias1993signal, sigl1994introduction}. Unlike the power spectrum, which is a second-order statistic reflecting only the distribution of power across frequencies, the bispectrum provides deeper insights by analyzing interactions between different frequency components \cite{nikias1993signal}. It is particularly effective in uncovering non-linear characteristics and harmonics in complex signals, making it a valuable method in advanced signal processing applications \cite{sigl1994introduction, kovach2018bispectrum}.

\begin{figure}[!t]
\centering
\includegraphics[width=0.98\textwidth]{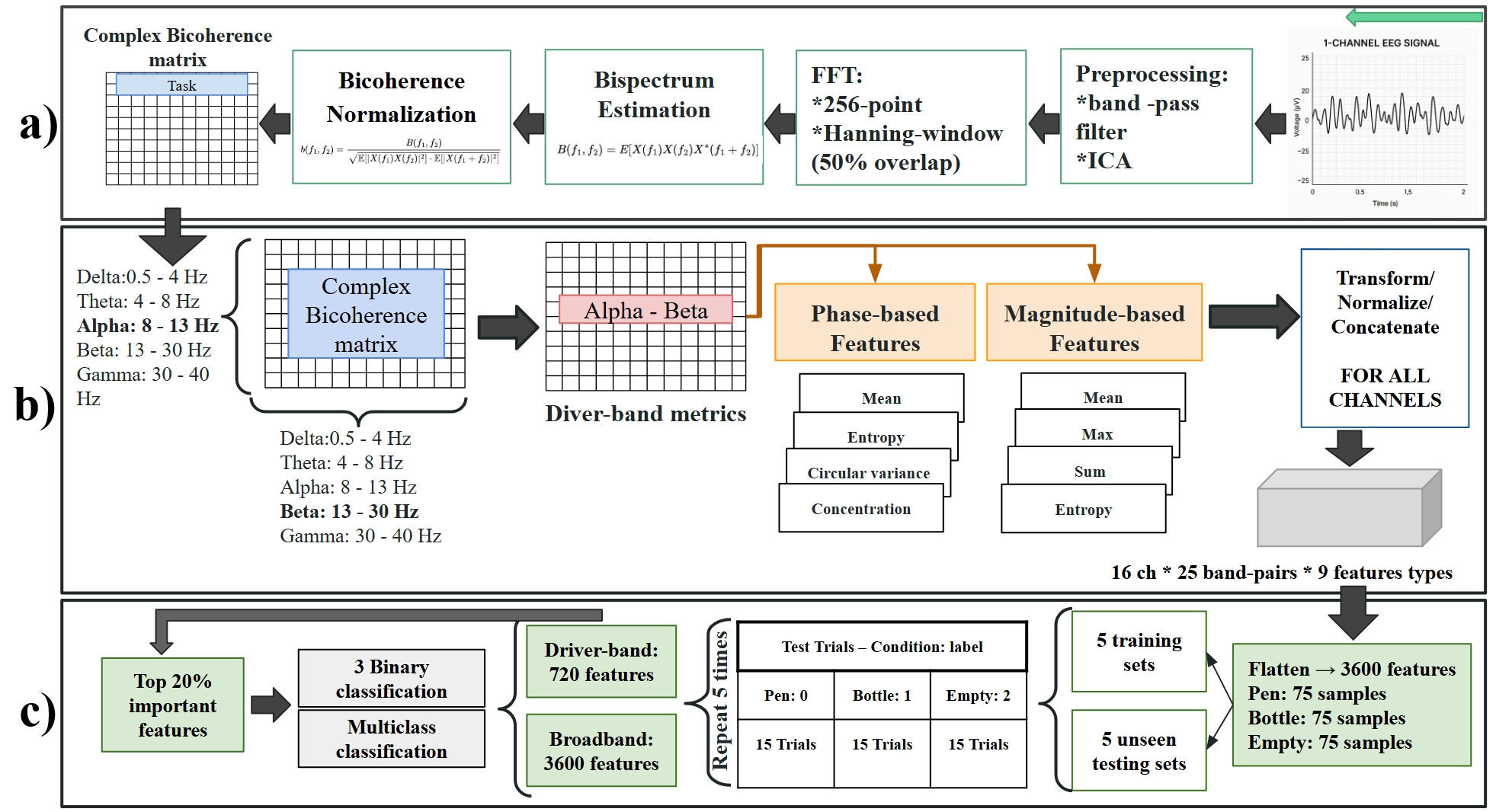}
\caption{Overview of the cross-frequency bispectral analysis and classification pipeline. (a) EEG preprocessing and bispectrum computation. (b) Derivation of cross-frequency driver–band matrices and extraction of phase- and magnitude-based bispectral features across all channels and band-pairs. (c) Feature selection and multi-task classification. }
\label{pipline}
\end{figure}

Building upon this, the cross-bispectrum extends the bispectrum concept to examine phase coupling and non-linear interactions between two different signals \cite{sigl1994introduction, chella2017non}. This analysis reveals coherence and shared phase relationships across systems, offering a deeper understanding of inter-signal dynamics and interactions, such as neural connectivity or communication between brain regions \cite{chella2017non, bartz2020beyond}. 

For a continuous signal \(x(t)\), the bispectrum \(B(f_1, f_2)\) is mathematically represented as \cite{nikias1993signal, sigl1994introduction}:

\begin{equation}
    B(f_1, f_2) =E[X(f_1) X(f_2) X^*(f_1 + f_2)]
\end{equation}

\noindent where \(X(f)\) is the Fourier transform of \(x(t)\), \(X^*(f)\) is its complex conjugate, and \(f_1\) and \(f_2\) are frequency components. $E[\cdot]$ represents expectation over multiple epochs. The bispectrum measures phase synchrony when the phases of \(X(f_1)\) and  \(X(f_2)\) sum to the phase of \(X(f_1+f_2)\), indicating nonlinear interactions \cite{nikias1993signal, sigl1994introduction, kovach2018bispectrum}.  In this study, the bispectrum was computed using a 256-point Fast Fourier Transform (FFT) with Hanning windows and 50\% overlap for time segments. Additional methodological details are provided in the Supplementary Material.

\subsection{Bicoherence}

Bicoherence is a normalized form of the bispectrum that retains both magnitude and phase information, providing a more comprehensive characterization of quadratic phase coupling between frequency components \cite{sigl1994introduction}. It is defined as:

\begin{equation}
    b(f_1, f_2) = \frac{B(f_1, f_2)}{\sqrt{\mathbb{E}[|X(f_1)X(f_2)|^2] \cdot \mathbb{E}[|X(f_1+f_2)|^2]}}
\end{equation}

Here, \(B(f_1, f_2)\) denotes the bispectrum, which quantifies the interaction among the frequencies \(f_1\), \(f_2\), and their sum \(f_1 + f_2\). The denominator normalizes the bispectrum, ensuring that the complex bicoherence is dimensionless and bounded in magnitude within the range [0, 1] \cite{sigl1994introduction}. Unlike squared bicoherence, the complex form preserves phase information, enabling the analysis of both the strength and directionality of phase coupling \cite{kovach2018bispectrum, nikias1993signal}. A magnitude close to 1 indicates strong phase synchronization, while the phase angle of \(b(f_1, f_2)\) provides insight into the relative phase relationships among the interacting frequency components \cite{sigl1994introduction, kovach2018bispectrum}. Figure \ref{pipline}-a shows the process of this computation for a given EEG channel. 

\subsection{Feature Extraction}
To systematically characterize the complex-valued bicoherence matrices, we extracted a set of quantitative features that capture both amplitude- and phase-related properties of nonlinear cross-frequency interactions. These features serve multiple analytical purposes. First, they enable classification between experimental conditions. Second, they support neurophysiologically meaningful interpretation of nonlinear phase coupling within and across EEG channels. Third, they facilitate statistical comparisons across tasks and time windows, revealing condition-specific variations in coupling strength and structure. Finally, they help identify dominant spatial and spectral signatures associated with motor planning and execution, highlighting the contribution of specific oscillatory dynamics to behavior.

Magnitude-based features include the mean, maximum, and sum of bicoherence values, along with the entropy of magnitudes, collectively quantifying the strength and spatial distribution of coupling across the frequency domain. Phase-based features include the mean phase angle, phase entropy, circular variance, and phase concentration (R), which together describe the directionality, variability, and consistency of phase relationships (Figure \ref{pipline}-b). All features served as the foundation for downstream classification, statistical analysis, and mechanistic interpretation. Further descriptions of the extracted features are provided in the Supplementary Material.

\subsection{Cross-Frequency Band-Pair Segmentation}

To investigate the spectral specificity of nonlinear cross-frequency coupling, we segmented each bicoherence matrix into predefined frequency band-pair regions based on canonical EEG rhythms. Specifically, we defined five standard frequency bands: delta (\( \delta \):, 1-4~Hz), theta (\( \theta \): 4–8 Hz), alpha (\( \alpha \): 8–13 Hz), beta (\( \beta \): 13–30 Hz), and gamma (\( \gamma \): 30–40 Hz). By pairing each band with every other—including itself—we constructed a total of 25 unique band-pair combinations (5 × 5). For each of these 25 pairs, we extracted the corresponding submatrix from the full bicoherence matrix and computed the same set of features described earlier.

Each element in the bicoherence matrix, \( b(f_1, f_2) \), captures the quadratic phase coupling among the frequency components \( f_1 \), \( f_2 \), and their sum \( f_1 + f_2 \). In this context, \( f_1 \) is often referred to as the \textit{modulating} or \textit{driver} frequency, and \( f_2 \) as the \textit{modulated} or \textit{responder} frequency. Segmenting the matrix into band-pairs based on this distinction allows us to investigate directional and asymmetric cross-frequency interactions. For example, the alpha–beta band-pair refers to the subset of the matrix where \( f_1 \) belongs to the alpha range (8–13 Hz) and \( f_2 \) to the beta range (13–30 Hz), thereby capturing how alpha oscillations may modulate or interact with beta-range dynamics (Figure \ref{pipline}-b).

This band-pair segmentation framework enables a fine-grained analysis of nonlinear interactions across the EEG spectrum, allowing us to isolate which specific spectral combinations—such as theta–gamma or alpha–beta—are differentially engaged across experimental conditions or behavioral states. By applying our feature extraction pipeline within each of these 25 subregions, we generated a rich and structured set of features that encode the strength and structure of cross-frequency interactions with both spatial and spectral resolution.

\subsubsection{Feature Matrix Construction and Preprocessing}

To construct a structured and interpretable representation of nonlinear cross-frequency coupling, we first computed the complex bicoherence matrix for each EEG channel, trial, and task stage. Each trial was segmented into two distinct 3-second windows corresponding to the planning and execution (reach-to-grasp) stages. For each trial and channel, we extracted the full bicoherence matrix and segmented it into 25 predefined frequency band-pair subregions. From each band-pair submatrix, we then computed eight quantitative features that capture both amplitude- and phase-based properties of the coupling, as described earlier. This process yielded a feature tensor of size \((75, 16, 25, 8)\) for each condition and stage, corresponding to 75 trials, 16 channels, 25 band-pairs, and 8 features.

To enable meaningful visualization and comparison of the feature space, we expanded the feature set to nine dimensions by decomposing the circular \texttt{mean\_phase} feature into its sine and cosine components. This transformation allowed us to preserve angular information while enabling linear analysis. Additionally, a logarithmic transformation was applied to the \texttt{sum\_bicoh} feature to reduce skewness. All features were then independently z-scored across trials, channels, and band-pairs to ensure comparability across scales and conditions. The final standardized feature tensor had shape \((75, 16, 25, 9)\) for each experimental condition and stage, resulting in six such tensors overall (three conditions \(\times\) two stages). Figure \ref{pipline}-b shows the complete procedure of band-pair segmentation, feature extraction, and construction.

\subsection{Feature Space Visualization and Discriminative Trends}

To explore the structure and interpretability of the extracted feature space, we conducted a series of visualizations aimed at identifying condition- or stage-specific patterns. First, we plotted feature-wise heatmaps, which allowed us to qualitatively assess which combinations of features and EEG channels may be sensitive to behavioral state, and to identify trends that could serve as candidate neural biomarkers of motor planning or execution.

To further explore discriminative patterns, we visualized classification accuracy trends across three binary tasks (Pen vs. Bottle, Pen vs. Empty, Bottle vs. Empty) and one multiclass task (Pen vs. Bottle vs. Empty), evaluated independently in both task stages and across two feature configurations (full feature set vs. single-band driver set). These plots provided a high-level view of how frequency-specific features and task stage influence classification behavior, motivating further statistical validation.

\subsection{Classification Pipeline and Feature Selection}

To evaluate the discriminative power of the extracted features, we employed a stratified 5-fold cross-validation approach for all classification tasks, allowing us to construct five unique train-test splits with non-overlapping test sets. Classification was performed separately for three binary tasks, as well as for a three-class classification task using Rnadom Forest. All tasks were evaluated independently within each stage (planning and execution), and under two feature scenarios: (1) using the full set of features across all band-pairs (3600 features per trial), and (2) using features derived from a single driver frequency band (720 features per trial). This structure resulted in a comprehensive analysis of classification performance across 4 tasks × 2 stages × 2 feature configurations.

Following initial classification, we estimated feature importance using a permutation-based approach applied to the training folds of a Random Forest model. For each feature, its contribution to classification accuracy was quantified by measuring the decrease in performance when the feature values were randomly shuffled, averaged across folds. This analysis allowed us to identify the most informative features with respect to the training data in each task and scenario. We then re-ran the classification pipeline using only the top-ranked features to assess whether a reduced feature set could preserve or improve discriminative performance. This secondary analysis enabled us to evaluate the robustness and generalizability of performance trends under dimensionality reduction and highlighted the potential for compact feature subsets in cross-frequency decoding (Figure \ref{pipline}-c). 

\subsection{Statistical Analysis}
A central objective of this study was to determine whether differences in 
classification performance across stages (Planning vs.\ execution), driver 
frequency bands (\(\delta, \theta, \alpha, \beta, \gamma\)), and tasks were 
statistically reliable rather than attributable to random variation \cite{button2013power}. 
Because cross-validated training accuracies provide multiple paired 
observations per subject and condition, they serve as a stable basis for 
within-subject statistical testing \cite{varoquaux2017assessing}. Accordingly, all statistical analyses described below were performed exclusively on the training 
accuracies obtained from the cross-validation procedure. The goal of these 
analysis is to quantify whether specific experimental factors (e.g., stage or 
frequency band) consistently influence model performance across subjects.

Each participant contributed a structured set of training accuracy values 
across: (1) two stages (Planning, execution), (2) five driver frequency bands,
and (3) four classification tasks. This repeated-measures structure allows
paired statistical testing while controlling for between-subject variability.

\subsubsection{Normality Assessment}
Before selecting appropriate statistical tests, the distribution of accuracy 
differences (e.g., execution $-$ Planning) was assessed using the 
Shapiro--Wilk test, which evaluates whether the sample distribution 
significantly deviates from a normal distribution. The Shapiro--Wilk test is
well-suited for small samples (e.g., \(N=10\) participants) and is widely 
recommended for neuroscience and biomedical datasets \cite{shaphiro1965analysis}.

When normality was satisfied (\(p > 0.05\)), parametric paired-samples 
\(t\)-tests were used. If the normality assumption was violated 
(\(p \leq 0.05\)), the Wilcoxon signed-rank test---a non-parametric 
alternative that does not assume Gaussianity---was applied. This decision 
ensures that statistical conclusions remain valid in the presence of 
non-normal or skewed distributions, which commonly arise in accuracy data \cite{wilcoxon1992individual}.

\subsubsection{Stage-Dependent Comparisons (Planning vs.\ Execution)}
To assess whether decoding performance differed between the Planning and 
execution stages, paired statistical comparisons were conducted for each 
driver band and task. Paired tests were chosen because the same subjects 
contributed accuracy values for both stages, allowing each participant to 
serve as their own control. This design isolates changes due to experimental 
effects rather than inter-individual differences.

When normality was satisfied, paired \(t\)-tests were used to evaluate 
mean differences. Effect sizes for paired \(t\)-tests were quantified using 
Cohen’s \(d\), where values of 0.2, 0.5, and 0.8 typically denote small, 
medium, and large effects, respectively \cite{cohen2013statistical}. For non-normal differences, the Wilcoxon signed-rank test was used, and 
effect sizes were expressed as rank-biserial correlations (\(r\)). This effect 
size measures how consistently one condition outperforms another and ranges 
from 0 (no difference) to 1 (complete separation).

Because five driver bands were tested within each task, resulting in five 
parallel comparisons, all \(p\)-values were corrected using the 
False Discovery Rate (FDR; Benjamini--Hochberg) procedure to control for 
false positives arising from multiple comparisons \cite{benjamini1995controlling}.

\subsubsection{Driver-Band Effects Across \(\delta\)--\(\gamma\)}
To evaluate whether classification performance differed across the five 
driver frequency bands within each task and stage, a non-parametric 
repeated-measures analysis was conducted using the Friedman test. The 
Friedman test is the non-parametric analogue of repeated-measures ANOVA 
and is appropriate because (1) the dataset contains repeated measurements 
from the same participants and (2) band-specific accuracies may not satisfy 
normality assumptions.

The Friedman test quantifies whether at least one frequency band produces 
systematically different accuracies across subjects. Its effect size, 
Kendall’s \(W\), measures the degree of agreement or consistency among the 
rankings of the five bands. Values close to 0 indicate weak or no 
band-dependent effects, whereas values approaching 1 indicate strong
agreement across subjects.

When the Friedman test indicated significant differences, post-hoc 
pairwise comparisons between driver bands were performed using the Wilcoxon 
signed-rank test. Post-hoc \(p\)-values were corrected using FDR to maintain 
statistical validity following multiple comparisons. Effect sizes for 
pairwise tests were again reported as rank-biserial correlations \cite{benjamini1995controlling, wilcoxon1992individual}.

\subsubsection{Exploratory Feature-Level Within-Subject Analysis}
In addition to accuracy-based statistical inference, an exploratory feature-level analysis was performed to examine stage-dependent modulation of individual bispectral features. Bispectral feature matrices for each subject and condition for the planning and execution stages were used. Feature values were aggregated across trials at the subject level, and paired within-subject statistical testing was conducted to compare planning versus execution for each individual feature (channel × band-pair × feature type).

Due to the high dimensionality of the feature space (3600 features) and the limited number of subjects, this analysis was considered exploratory and aimed at identifying focal, subject-consistent stage transitions rather than establishing widespread statistical significance. FDR correction was applied across all tested features to control for multiple comparisons. Results from this analysis were interpreted in conjunction with classification performance and permutation-based feature importance, and were not used for model selection or parameter tuning \cite{benjamini1995controlling, kriegeskorte2009circular}.

\section{Results}
This section presents a comprehensive evaluation of cross-frequency bispectral EEG features to characterize motor planning and execution during reach-to-grasp tasks. We first present descriptive visualizations of extracted bispectral features to illustrate their spatial and spectral distributions across task conditions and stages. Classification performance is then evaluated across multiple task contrasts, motor stages, and driver frequency bands, followed by within-subject statistical analyses assessing stage- and band-dependent effects. Generalization performance on unseen test data and the impact of permutation-based feature selection are subsequently examined. Finally, feature-importance analyses are used to summarize the relative contributions of EEG channels, frequency band pairs, and bispectral feature types.
\begin{figure*}[!t]
\centering
\includegraphics[width=0.7\textwidth]{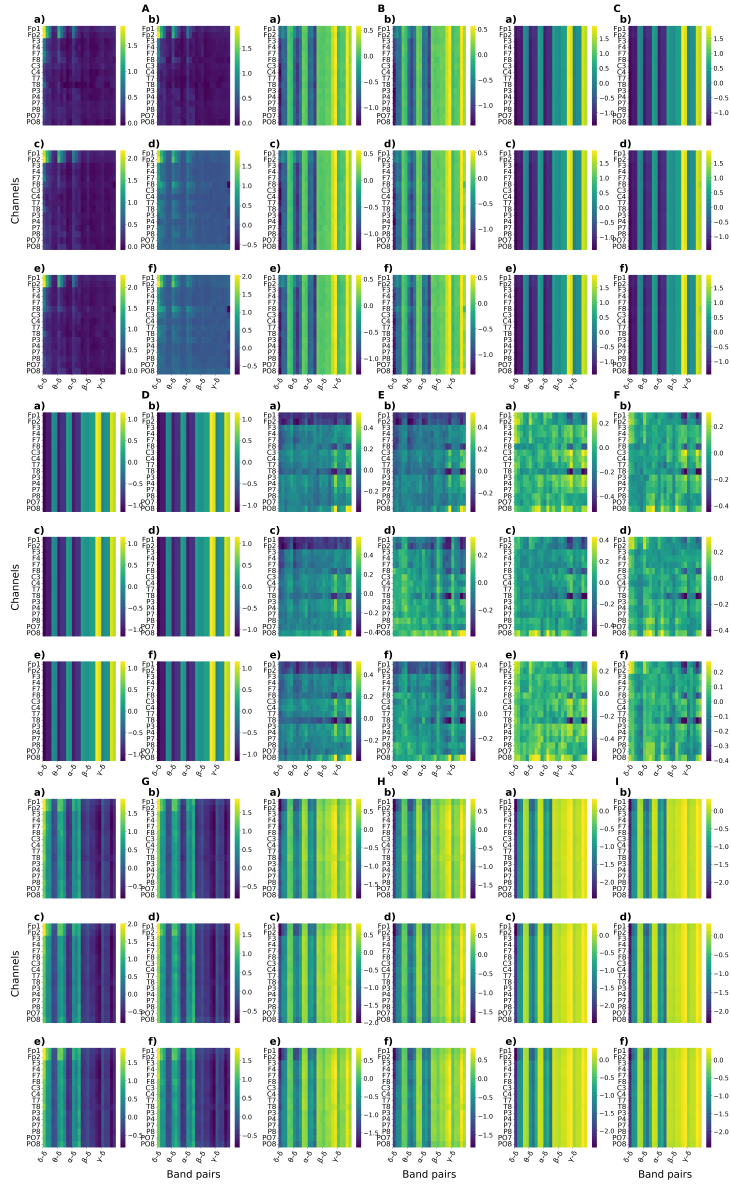}
\caption{Subject-averaged feature maps across channels and frequency band pairs.  Each large panel (A-I) corresponds to one feature: A: mean, B: max, C: sum, D: entropy, E: sine of the biphase, F: cosine of the biphase, G: phase-coupling concentration $R$, H: circular variance, and I: phase entropy. Within each feature block, six subplots are for each task condition: NM, bottle, and pen in planning (a, c, e), and in execution (b, d, f).}
\label{visual}
\end{figure*}

\subsection{Descriptive Visualization of Feature Space}
Figure \ref{visual} illustrates the group-level distribution of all bispectral and phase-based features across channels and band pairs for each task condition. As seen in the averaged maps, the overall spatial–spectral patterns remain relatively stable across conditions, with no large, visually striking differences between planning and execution or between object types. The bispectral amplitude features tend to show the strongest structured patterns, particularly prominent low-frequency driver interactions projecting into higher-frequency bands, whereas the phase-based measures exhibit more diffuse and channel-dependent variability. Although these patterns suggest that condition-related differences are subtle at the group level, small but systematic variations are still present across feature families and task stages. This supports the notion that discriminability in later classification results likely arises from finer-scale, subject-specific deviations rather than from large global shifts visible in averaged feature maps.

\subsubsection{Exploratory Feature-Level Within-Subject Analysis}
An exploratory feature-level within-subject analysis was conducted to identify individual bispectral features exhibiting consistent phase-dependent modulation between the planning and execution stages. Across the full 3600-dimensional feature space, only a very limited number of features survived global false discovery rate (FDR) correction, reflecting the conservative nature of univariate testing in a high-dimensional setting. Notably, two features in the precision grasp (\emph{Pen}) condition remained significant after correction (minimum $p_{\mathrm{FDR}} = 0.0136$). Both features corresponded to the sine of the bispectral phase (\texttt{sin\_phase}) at the central electrode \texttt{Cz}, specifically within theta--beta and beta--theta band-pair interactions. These features showed a consistent negative shift from planning to execution across subjects, indicating a systematic reorganization of cross-frequency phase relationships during execution rather than a change in coupling magnitude. No features survived global FDR correction for the \emph{Bottle} or no-execution conditions, although numerous features exhibited uncorrected phase-dependent differences, suggesting more distributed and heterogeneous modulation patterns. Overall, these findings indicate that execution-related neural dynamics during precision grasping are characterized by focal, subject-consistent phase reorganization at central motor regions, whereas other conditions rely on broader multivariate feature patterns that are not captured by univariate significance alone.

\subsection{Classification Performance Across Tasks, Stages, and Driver Bands}

\begin{figure*}[!t]
\centering
\includegraphics[width=0.98\textwidth]{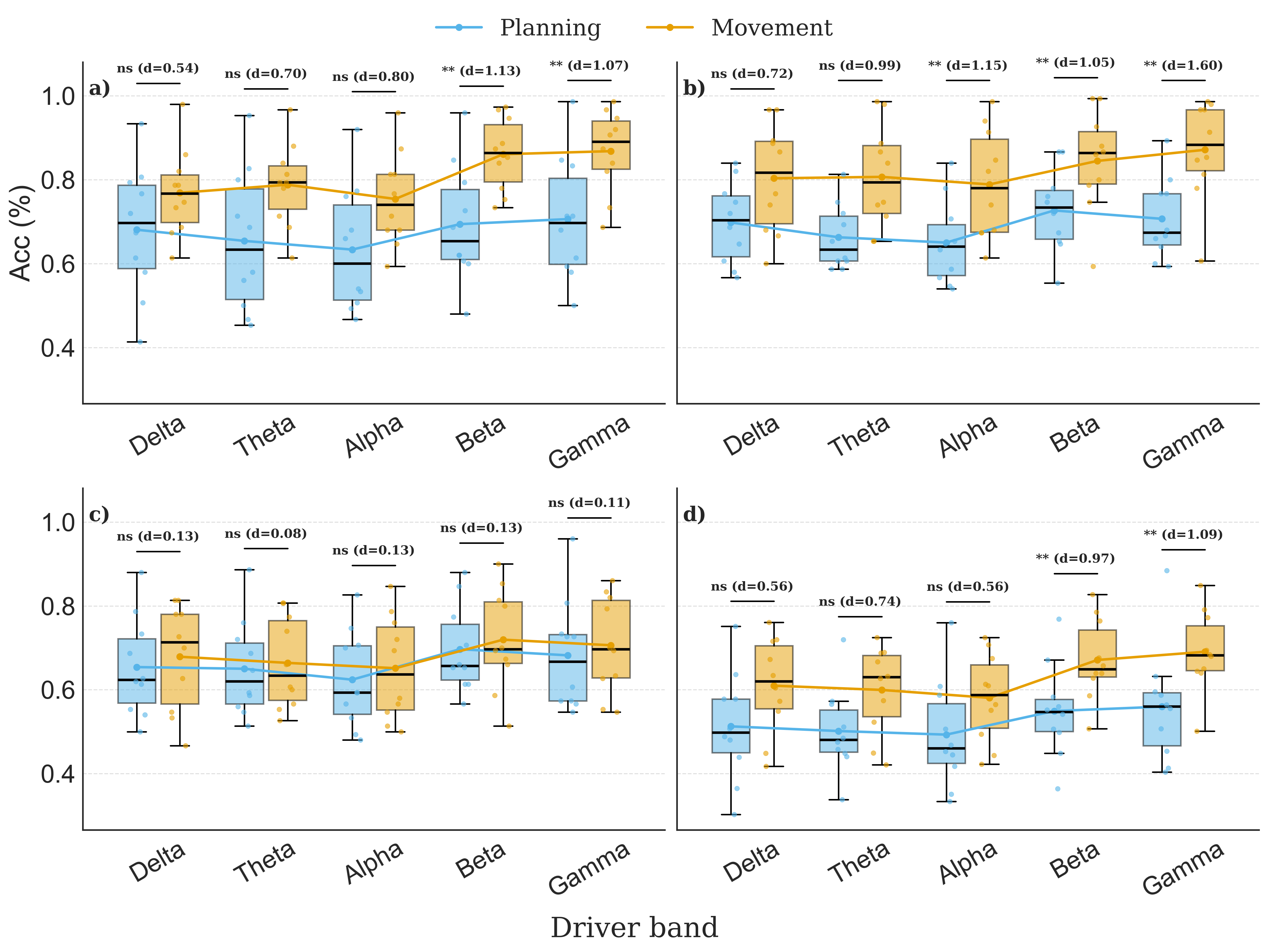}
\caption{Classification accuracies (10-fold cross-validation) across frequency bands ($\delta$, $\theta$, $\alpha$, $\beta$, $\gamma$) for Planning (blue) and execution (orange) stages in four task settings: (a) Power grasp vs.\ No-execution, (b) Precision grasp vs.\ No-execution, (c) Power vs.\ Precision grasp, and (d) Multiclass (Power, Precision, No-execution). Each boxplot summarizes subject-level accuracies ($N=10$). Horizontal bars above each band indicate paired comparisons between stages, with statistical outcomes corrected for multiple comparisons (FDR): \textit{ns} = non-significant, and ** is significant $p < 0.05$. Effect sizes (Cohen’s $d$) are shown in parentheses.}

\label{training_trend}
\end{figure*}

\subsubsection{Stage-dependent Differences Across Driver Bands}
% To evaluate whether decoding performance differed between Planning and execution stages, we conducted paired within-subject comparisons across driver frequency bands (e.g., $\gamma$--$\delta$, $\gamma$--$\theta$, …, $\gamma$--$\gamma$). For each subject, classification accuracies were matched between stages within the same task and driver band. Normality of stage differences (execution $-$ Planning) was assessed using the Shapiro–Wilk test, and all comparisons satisfied normality assumptions ($p > 0.05$). Therefore, paired-samples \textit{t}-tests were applied throughout. Effect sizes were quantified using Cohen’s $d$. To correct for multiple comparisons across tasks and bands, $p$-values were adjusted using the false discovery rate (FDR, Benjamini–Hochberg).

Figure~\ref{training_trend} summarizes subject-level classification accuracies (10-fold cross-validation) for Planning (blue) and execution (orange) stages across driver bands in four task conditions: (a) Power vs. No-execution, (b) Precision vs. No-execution, (c) Power vs. Precision, and (d) Multiclass (Power, Precision, No-execution). Horizontal bars above each driver band indicate paired comparisons, with annotations denoting statistical outcomes after FDR correction (\textit{ns} = non-significant; $p < 0.05$ (**)). Effect sizes are shown in parentheses.

In the \textbf{Power vs. No-execution} condition (Fig.~\ref{training_trend}-a), accuracies during Planning ranged from 0.63 (alpha) to 0.73 (beta), while execution improved to 0.75 (alpha) and 0.84 (beta). The difference was most pronounced in the $\beta$ and $\gamma$ driver bands, where execution significantly outperformed Planning ($p_{FDR} < 0.05$), with large effect sizes ($d = 1.13$ for $\beta$, $d = 1.07$ for $\gamma$). In lower-frequency driver bands (delta, theta), execution also trended higher (mean increases of 0.07–0.12), but these did not survive FDR correction.

For \textbf{Precision vs. No-execution} (Fig.~\ref{training_trend}-b), the advantage of execution was even more robust. Planning accuracies averaged 0.65–0.71 across bands, whereas execution improved to 0.79–0.87, yielding significant differences in the $\alpha$ ($d = 1.15$), $\beta$ ($d = 1.05$), and $\gamma$ ($d = 1.60$) driver bands ($p_{FDR} < 0.05$). The effect sizes here were among the largest observed in the dataset, highlighting the strong contribution of higher-frequency bands to discriminating Precision grasps from baseline during execution.

In contrast, \textbf{Power vs. Precision} (Fig.~\ref{training_trend}-c) did not show reliable stage-related effects. Accuracy values were relatively balanced between Planning (0.62–0.70) and execution (0.65–0.72), and none of the driver bands reached significance after FDR correction. Effect sizes were small ($d \leq 0.13$), suggesting that grasp-type distinctions are encoded similarly during planning and execution.

Finally, in the \textbf{Multiclass condition} (Fig.~\ref{training_trend}-d), Planning accuracies averaged 0.49–0.60 across bands, while execution improved to 0.57–0.69. Here, significant differences emerged in the $\beta$ ($d = 0.97$) and $\gamma$ ($d = 1.09$) driver bands ($p_{FDR} < 0.05$), again indicating a strong execution-stage enhancement in higher-frequency oscillatory interactions.

Taken together, these results reveal a consistent trend: execution stage decoding accuracies exceeded Planning across nearly all driver bands and tasks, with the most robust improvements localized to $\beta$ and $\gamma$ driver bands. While task-vs.-baseline contrasts (Power vs. No-execution, Precision vs. No-execution, Multiclass) showed clear stage dependence, the grasp-type contrast (Power vs. Precision) remained stable across stages. These findings suggest that execution execution engages stronger cross-frequency coupling patterns than planning, particularly in high-frequency driven bands.

\begin{table}[!t]
\centering
\renewcommand{\arraystretch}{1.4}  % makes rows taller
\caption{Friedman test results for band-level differences in classification accuracy across tasks and stages (Plan vs. Ex., where Ex. denotes the execution stage). Kendall’s W represents effect size, and significant FDR-corrected post-hoc comparisons are listed with rank-biserial correlations.}
\begin{tabular}{
    >{\centering\arraybackslash}m{1cm}
    >{\centering\arraybackslash}m{2cm}
    >{\centering\arraybackslash}m{1.5cm}
    >{\centering\arraybackslash}m{1.5cm}
    >{\centering\arraybackslash}m{5cm}
}
% {p{1cm} p{2cm} p{1.5cm} p{1.5cm} p{5cm}}
\hline
stage & Task & Friedman $p$ & Kendall's $W$ & Significant post-hoc (FDR) \\
\hline
Ex.*& Power vs. NM& 0.0003 & 0.53 & $\beta > \delta$ ($r=0.60$); $\gamma > \delta$ ($r=0.80$); $\gamma > \theta$ ($r=0.80$); $\gamma > \alpha$ ($r=0.80$) \\
Ex.& Precision vs. NM& 0.0469 & 0.24 & None \\
Ex.& Power vs. Precision& 0.0015 & 0.43 & $\beta > \theta$ ($r=0.80$); $\gamma > \theta$ ($r=0.60$); $\gamma > \alpha$ ($r=0.80$) \\
Ex.& Multi    & 0.0001 & 0.57 & $\gamma > \delta$ ($r=0.60$); $\beta > \theta$ ($r=0.80$); $\gamma > \alpha$ ($r=1.00$); $\gamma > \beta$ ($r=1.00$) \\ \hline
Plan & Power vs. NM& 0.0343 & 0.26 & $\gamma > \theta$ ($r=0.80$); $\gamma > \alpha$ ($r=0.80$) \\
Plan & Precision vs. NM& 0.0432 & 0.24 & None \\
Plan & Power vs. Precision& 0.0825 & 0.21 & None \\
Plan & Multi    & 0.0320 & 0.26 & None \\
\hline
\end{tabular}
\label{tab:band_effects}
\end{table}

\subsubsection{Driver Frequency Bands Effects Analysis}
To assess whether classification performance differed across frequency bands within each task and stage, we conducted a non-parametric repeated-measures analysis (Table \ref{tab:band_effects}). 

% Specifically, a Friedman test was applied across the five driver bands ($\delta$, $\theta$, $\alpha$, $\beta$, $\gamma$) for each subject, task (Power vs.\ No-execution, Precision vs. No-execution, Power vs.\ Precision, Multiclass), and stage (Planning, execution). The Friedman test was chosen as the non-parametric alternative to repeated-measures ANOVA because it does not assume normality and is appropriate for within-subject designs with multiple conditions. 

Kendall’s $W$ was computed as an effect size, quantifying the degree of concordance across bands ($W=0$ indicates no agreement; $W=1$ indicates perfect agreement). When overall band effects were significant, post-hoc pairwise comparisons were performed, with effect sizes reported as rank-biserial correlations ($r$). All post-hoc $p$-values were corrected for multiple comparisons using FDR.

The  analysis revealed distinct patterns across stages. During the Planning stage, band effects were weak to moderate (Kendall’s $W=0.21$–$0.26$), and only the Power vs.\ No-execution task showed significant differences across bands ($p=0.0343$). Post-hoc comparisons indicated that $\gamma$-driver features outperformed $\theta$ and $\alpha$ features (both $r=0.80$). However, in all other Planning tasks, no pairwise differences survived FDR correction, suggesting broadly comparable classification performance across bands.

In contrast, the execution stage exhibited robust and consistent band effects (Friedman $p<0.01$, Kendall’s $W=0.37$–$0.57$). Post-hoc tests revealed that $\gamma$-driver features consistently outperformed other bands, with large effect sizes ($r=0.60$–$1.00$). For example, in the Power vs.\ No-execution task, $\gamma$ exceeded $\delta$, $\theta$, and $\alpha$, and $\beta$ also outperformed $\delta$ ($r=0.60$). In the Precision vs.\ No-execution task, no pairwise differences reached significance, although the global effect was marginal ($p=0.0469$). For Power vs.\ Precision, $\gamma$ surpassed $\theta$ ($r=0.60$) and $\alpha$ ($r=0.80$), while $\beta$ outperformed $\theta$ ($r=0.80$). In the Multiclass setting, $\gamma$ again outperformed $\delta$, $\alpha$, and $\beta$, while $\beta$ exceeded $\theta$. Taken together, these results demonstrate that while Planning accuracy was relatively homogeneous across bands, execution accuracy was strongly band-dependent, with $\gamma$ consistently providing the most discriminative feature set. This suggests that bispectral features anchored in higher-frequency activity capture the most robust task-related information during motor execution.

\begin{table}
    \centering
    \renewcommand{\arraystretch}{1.3}  % makes rows taller
    \begin{tabular}{
    >{\centering\arraybackslash}m{0.7cm}
    >{\centering\arraybackslash}m{2.6cm}
    >{\centering\arraybackslash}m{3.2cm}
    >{\centering\arraybackslash}m{3cm}
    >{\centering\arraybackslash}m{2.5cm}
    } \hline
         &  \multicolumn{4}{c}{Planning}\\ \hline
         Driver Band &  0 vs. 1&  0 vs. 2&  2 vs. 1& 0 vs. 1 vs. 2\\ \hline
         \( \delta \) &  $66.46\pm12.86$ &  $61.26\pm12.87$&  $66.66\pm11.33$& $49.82\pm12.14$ \\
         \( \theta \) &  $61.53\pm15.54$&  $64.06\pm11.77$&  $65.46\pm9.46$& $48.80\pm12.09$ \\
         \( \alpha \) &  $61.53\pm14.69$&  $59.66\pm13.02$&  $62.93\pm9.63$& $47.22\pm11.79$ \\
         \( \beta \) &  $67.00\pm12.21$&  $66.66\pm9.69$&  $68.40\pm8.68$& $51.73\pm10.91$ \\
         \( \gamma \) &  $67.46\pm14.28$&  $65.46\pm12.48$&  $67.00\pm8.77$& $53.06\pm12.70$ \\
         WB&  $69.20\pm14.38$&  $67.66\pm10.71$&  $70.40\pm10.91$& $56.22\pm13.32$ \\
 Top& $72.86\pm12.70$& $75.62\pm10.66$& $79.13\pm10.28$&-\\ \hline
 & \multicolumn{4}{c}{Execution}\\ \hline
 \( \delta \) 
& $73.80\pm12.02$& $64.60\pm10.85$& $80.13\pm13.26$&$59.06\pm8.62$ \\
 \( \theta \) 
& $74.13\pm12.64$& $64.40\pm10.04$& $78.00\pm12.63$&$58.17\pm9.93$ \\
 \( \alpha \) 
& $71.80\pm13.48$& $61.13\pm8.77$& $76.53\pm13.58$&$56.00\pm11.70$ \\
 \( \beta \) 
& $82.73\pm10.48$& $68.06\pm11.38$& $83.53\pm10.91$&$62.97\pm10.76$ \\
 \( \gamma \) 
& $83.60\pm11.73$& $66.00\pm10.31$& $85.33\pm12.66$&$65.02\pm9.24$ \\
 WB& $85.86\pm10.44$& $66.86\pm11.67$& $87.06\pm11.15$&$68.93\pm8.57$ \\
 Top& $83.33\pm5.90$& $78.19\pm11.44$& $82.33\pm5.63$&-\\ \hline
    \end{tabular}
    \caption{Classification accuracies across frequency bands, whole-band features, and selected top features for both Planning and execution stages. Values correspond to mean accuracies ($\pm$ standard deviation) averaged across subjects, for binary classification tasks as well as for the multiclass condition (Power/Bottle (C:0) vs.\ Precision/Pen (C: 2) vs.\ No-execution/Empty (C:1)). Rows correspond to spectral bands ($\delta$, $\theta$, $\alpha$, $\beta$, $\gamma$), whole-band features, and the subset of top-ranked features. }
    \label{tab:tableAcc}
\end{table}

\subsection{Generalization Performance on Unseen Test Sets}

Table~\ref{tab:tableAcc} summarizes the classification accuracies obtained on \textbf{unseen test sets}, averaged across ten subjects. For each subject, results were first averaged over five independent test sets and subsequently across subjects, ensuring robustness of the reported values. Results are expressed as mean accuracy ($\pm$ standard deviation). Importantly, the analyses are based on bispectrum-derived features, where each \emph{driver band} ($\delta$, $\theta$, $\alpha$, $\beta$, $\gamma$) contributed 720 features, yielding a total of 3600 features across all band-pair combinations. The ``Whole-band'' category corresponds to using all features pooled across driver bands (full 3600 features), while the ``Top features'' subset reflects features retained after permutation importance testing (score $>$ 0.00), thus representing the most informative subset.  

During the \textbf{Planning stage}, classification using individual driver bands yielded moderate accuracies. For example, $\delta$-band features reached $66.46 \pm 12.86\%$ (Power vs. NM), while $\gamma$-band features achieved $67.46 \pm 14.28\%$. Whole-band features, which pooled across all driver bands, increased performance to $69.20 \pm 14.38\%$ for Power vs. NM and $70.40 \pm 10.91\%$ for Precision vs. NM. Notably, applying permutation-based feature selection provided substantial gains: top features improved accuracies to $72.86 \pm 12.70\%$ for Power vs. NM, $75.62 \pm 10.66\%$ for Power vs. Precision, and $79.13 \pm 10.28\%$ for Precision vs. NM. This shows that carefully chosen subsets of features outperform both isolated driver bands and brute-force whole-band features. Multiclass decoding during Planning remained challenging, with accuracies around chance, reaching $56.22 \pm 13.32\%$ in the whole-band case.  

In the \textbf{execution stage}, unseen test accuracies were consistently higher across all conditions. Driver bands $\beta$ and $\gamma$ were particularly discriminative, reaching $82.73 \pm 10.48\%$ and $83.60 \pm 11.73\%$ respectively for Power vs. NM. The $\delta$ band also contributed strongly, achieving $80.13 \pm 13.26\%$ for Precision vs. NM. Whole-band features again provided an advantage, with accuracies up to $87.06 \pm 11.15\%$ (Precision vs. NM), the highest observed value in this table. The top feature subset yielded comparable performance (e.g., $83.33 \pm 5.90\%$ for Power vs. NM), confirming that feature selection preserves generalization while reducing dimensionality. Even for multiclass decoding, which is intrinsically more difficult, execution-stage accuracies remained above chance: $68.93 \pm 8.57\%$ with whole-band features and $65.02 \pm 9.24\%$ for the $\gamma$ band.  

Collectively, these results demonstrate that generalization to unseen test data is substantially stronger during execution compared to Planning, with higher-frequency driver bands ($\beta$, $\gamma$) and pooled whole-band features providing the strongest discriminative power. The performance boost observed when restricting to permutation-derived top features highlights the importance of feature relevance, particularly in the Planning stage where classification is more difficult. While multiclass tasks remain less accurate overall, the above-chance performance indicates that both driver-band-specific and whole-band bispectrum features contain meaningful neural signatures of task-related processes.

\subsection{Feature Importance and Reduced-Dimension Performance}

\begin{figure*}[!t]
\centering
\includegraphics[width=0.98\textwidth]{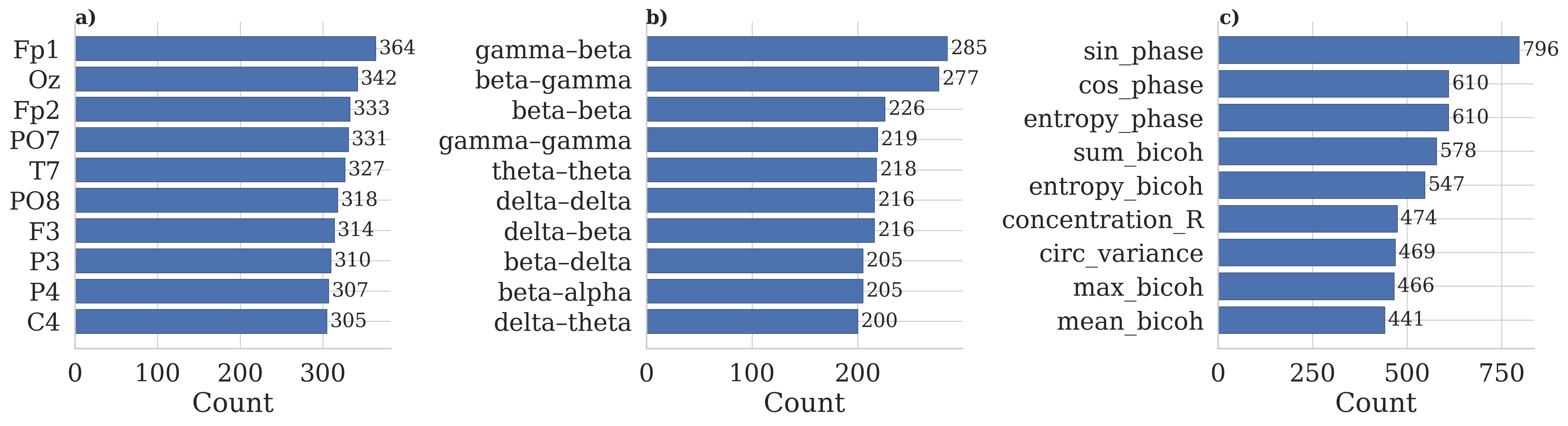}
\caption{Most frequently selected feature components across all subjects, conditions, and task phases.
(a) Top 10 EEG channels, (b) Top 10 band-pair interactions, and (c) Top bispectral feature types, based on their global frequency of selection across all classification models. Bar heights indicate the total number of times each component appeared among the top-ranked features. This global summary highlights the most consistently informative spatial (channel), spectral (band-pair), and statistical (feature type) domains in the bispectral feature space across subjects.}
\label{fig:selection_count}
\end{figure*}

\subsubsection{Global Feature Selection Profile Across Subjects and Phases}

To better understand which components of the bispectral feature space consistently contributed to classification, we performed a global summary analysis of the most frequently selected features across all subjects, conditions, and task phases. From the full feature pool of 3600 candidates (16 EEG channels $\times$ 25 band-pair interactions $\times$ 9 feature types), we applied permutation testing to evaluate feature importance and identify those with non-zero contribution scores. The bar plots in Fig.~\ref{fig:selection_count} show the top ten most frequently selected (a) EEG channels, (b) driver band-pair interactions, and (c) feature types, based on their aggregated frequency of selection across all classification runs. The heights of the bars indicate how many times each feature dimension appeared among the top-ranked subset after permutation testing, providing an overall measure of robustness and consistency.

The results highlight a distinct pattern across spatial, spectral, and statistical domains. Spatially, prefrontal (Fp1, Fp2) and occipital (Oz, PO7, PO8) regions were among the most frequently selected, suggesting strong involvement of both frontal control and visual processing areas. Spectrally, interactions involving higher-frequency drivers (e.g., gamma–beta, beta–gamma, and gamma–gamma) were dominant, followed by cross-frequency couplings between theta and delta bands, indicating that both local high-frequency and long-range low-frequency coordination contribute to task representation. At the feature-type level, phase-related measures (sine and cosine of phase, phase entropy) were the most robustly selected, followed by bispectral magnitude features (sum, entropy, max, mean). This global summary demonstrates that the most informative features emerge from a combination of prefrontal/occipital regions, gamma- and beta-driven interactions, and phase-based metrics.

\subsubsection{Cross-frequency Band–pair Contributions During Planning and Execution Phases }
\begin{figure*}[!t]
\centering
\includegraphics[width=0.98\textwidth]{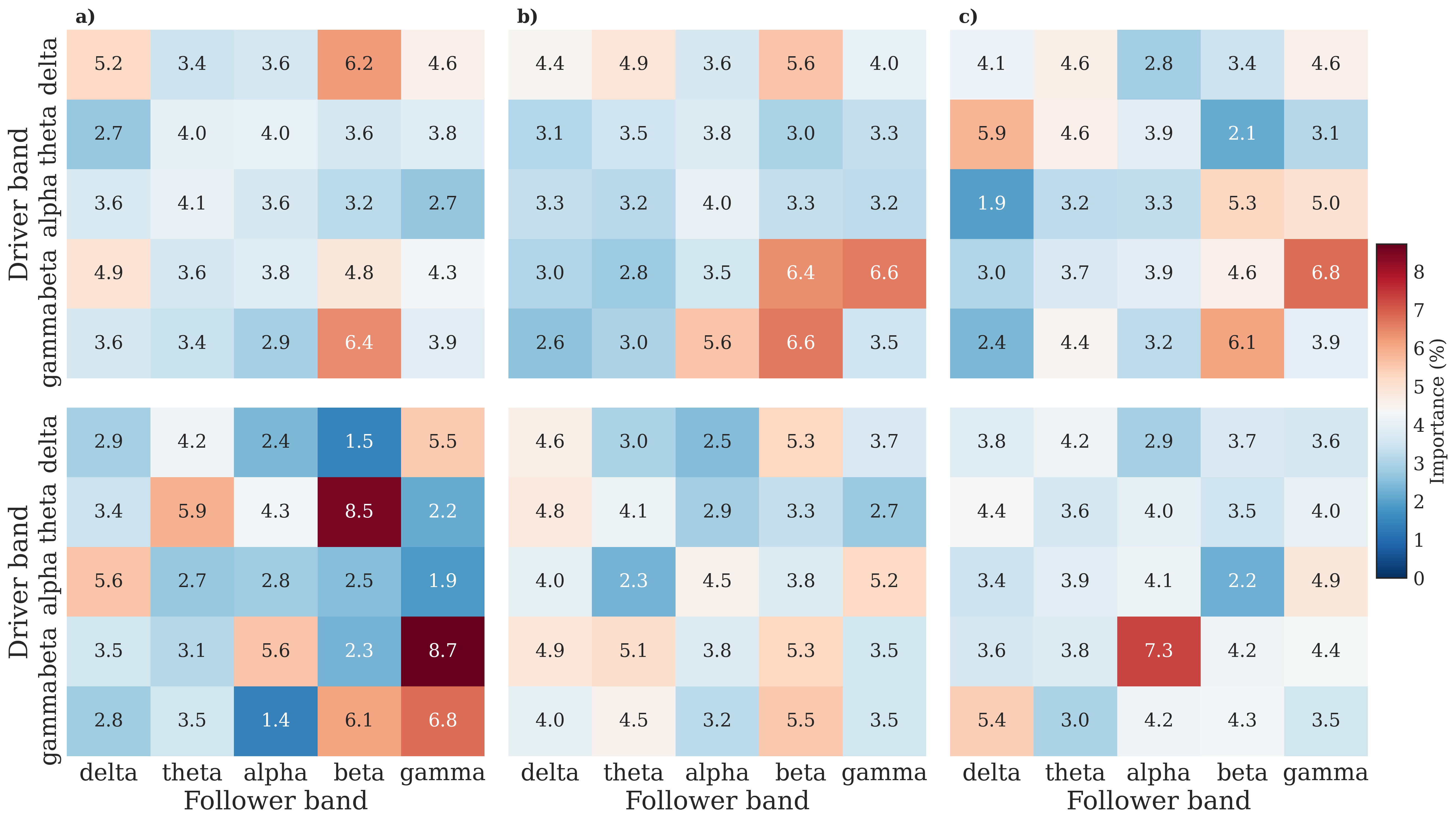}
 \caption{Group-level distribution of feature importance across frequency band-pairs 
    during planning (top row) and execution (bottom row) phases for three classification tasks:  (a) Pen vs. No-execution (NM), (b) Bottle vs. NM, and (c) Bottle vs. Pen. 
    Each 5$\times$5 heatmap shows the relative contribution (\%) of each driver--follower band-pair to the overall classification importance, averaged across subjects and normalized within each condition and phase. Rows correspond to driver bands (delta, theta, alpha, beta, gamma),  and columns correspond to follower bands. Warmer colors indicate higher contribution values. }
    \label{fig:bandpair_heatmaps}
\end{figure*}
Figure~\ref{fig:bandpair_heatmaps} illustrates the relative distribution of feature importance across driver--follower frequency band-pairs for the three classification tasks (Pen vs.\ NM, Bottle vs.\ NM, and Bottle vs.\ Pen) during planning and execution phases. To derive the values presented in Figure~\ref{fig:bandpair_heatmaps}, permutation-based feature importance scores were first obtained for each subject, condition, and phase. Features were grouped according to their frequency band-pair (e.g., beta--gamma, theta--alpha) regardless of the contributing channel or feature type. For each subject, the scores within a given band-pair were summed across all channels and feature types, yielding a total contribution of that band-pair. These values were then averaged across subjects and normalized within each condition and phase such that the band-pair contributions summed to 100\%. The resulting percentages, shown in each heatmap cell, therefore reflect the relative importance of driver--follower frequency interactions in contributing to classification performance. 

During the \textbf{planning phase} (Figure \ref{fig:bandpair_heatmaps}, top row), contributions were broadly distributed across multiple band-pairs, with values ranging between 2--7\%. Despite this distributed profile, several consistent trends emerged. Beta--gamma and gamma--beta interactions accounted for the strongest contributions in the Bottle vs.\ NM contrast ($\sim$6--7\%), whereas delta--beta and gamma--beta couplings were more prominent in the Pen vs.\ NM comparison ($\sim$6\%). For Bottle vs.\ Pen, theta--delta and beta--gamma interactions contributed relatively higher percentages ($\sim$6\%), suggesting a heterogeneous network of cross-frequency dynamics engaged during action planning.

In contrast, the \textbf{execution phase} (bottom row) revealed more focal patterns of contribution. In the Pen vs.\ NM condition, theta--beta and beta--gamma couplings exhibited the strongest peaks, exceeding 8\% of the total importance, alongside a notable contribution from alpha--delta interactions. For Bottle vs.\ NM, relative contributions were more moderate, though delta--beta and gamma--beta remained among the stronger pairs ($\sim$5\%). Importantly, the Bottle vs.\ Pen condition demonstrated a distinct hotspot at beta--alpha (7.3\%), reflecting differential cross-frequency coupling mechanisms underlying precision and power grasp execution.

Together, these results indicate that planning is supported by a distributed set of cross-frequency interactions spanning low- and high-frequency bands, whereas execution execution involves sharper and more condition-specific contributions, particularly within beta-centered couplings. These patterns highlight the functional specificity of cross-frequency interactions in discriminating between motor states and grasp types.

\section{Discussion}

\subsection{Summary of Key Findings and Contributions}
This study presents the first comprehensive application of cross-frequency bispectral EEG analysis to executed reach-to-grasp executions involving precision and power grips. Unlike prior work that focuses primarily on motor imagery and limited bispectral descriptors, our framework demonstrates that nonlinear cross-frequency interactions contain rich and discriminative signatures of both motor planning and execution execution. Four key contributions emerge: (1) extending bispectral analysis to ecologically valid grasp execution, (2) demonstrating the discriminative power of bispectrum-derived features for natural object-specific grasp decoding, (3) revealing phase-dependent insights into neural mechanisms underlying planning versus execution, and (4) providing a structured analysis of frequency band-pair interactions that clarifies how specific oscillatory couplings support grasp behavior. Together, these contributions establish a methodological and conceptual foundation for integrating higher-order spectral features into future neuroprosthetic and BCI systems.

\subsection{Nonlinear Cross-Frequency Dynamics in Planning and Execution}
A central finding is the consistent superiority of execution-phase decoding over the Planning phase across tasks and frequency bands (Figure \ref{training_trend}). Statistical analyses (Table \ref{tab:band_effects}) revealed that $\beta$- and $\gamma$-driver features exhibit particularly strong enhancements during execution, with large effect sizes in contrasts involving object grasping versus baseline. These observations align with established neurophysiology showing that beta desynchronization, gamma synchronization,
and their cross-frequency coordination intensify during active motor engagement and sensorimotor integration \cite{cheyne2013meg, jensen2007cross}. Together, these results suggest that execution execution is characterized by a more focal, high-frequency–anchored pattern of nonlinear neural coordination.

In contrast, power- versus precision-grasp decoding (Figure \ref{training_trend}) exhibited comparable performance across phases, indicating that grasp-type representations emerge early during planning and are preserved through execution. This suggests that grip selection engages premotor networks prior to execution initiation. This finding is consistent with the involvement of premotor and associative networks during execution preparation, which may support a more distributed, multi-frequency organization of neural activity before execution onset. Collectively, the results highlight a transition from distributed nonlinear coordination during planning to more spectrally specific coupling during execution, demonstrating that nonlinear cross-frequency coupling captures complementary preparatory and execution-related neural processes that traditional spectral-power analyses may overlook \cite{cisek2010neural, rabiee2025learningmultimodalaialgorithms}.

\subsection{Spectral Specificity of Discriminative Features}
Driver-band analyses revealed strong spectral specificity, with $\gamma$-driver features consistently outperforming other bands during execution, followed by $\beta$-driver features. These findings were reproduced across cross-validated training accuracies (Figure \ref{training_trend}), unseen test-set performance (Table \ref{tab:tableAcc}), and permutation-based feature selection (Figure \ref{fig:selection_count}). High-frequency bands likely reflect nonlinear interactions within fast sensorimotor circuits responsible for grasp force modulation, rapid feedback control, and corticomuscular coordination. 
During planning, band effects were weaker and more broadly distributed across low- and high-   frequency combinations, suggesting that motor preparation relies on a diffuse cross-frequency landscape, whereas execution sharpens into high-frequency– anchored coupling patterns. This transition underscores the dynamic reorganization of brain networks across the temporal evolution of a reach-to-grasp action.

\subsection{Spatial Patterns: Prefrontal and Occipital Contributions}
Global feature-selection results (Figure \ref{fig:selection_count}) indicated that prefrontal (Fp1, Fp2) and occipital (Oz, PO7, PO8) regions were the most consistently selected channels across tasks and phases. Prefrontal involvement reflects attention, cognitive control, and decision-related processes essential during planning. Occipital contributions highlight the role of visual processing and object identification, which remain relevant during both planning and early execution. These spatial patterns emphasize that reach-to-grasp behavior integrates distributed cognitive, sensorimotor, and visual systems, and that bispectral coupling effectively captures these coordinated dynamics.

\subsection{Band-Pair Contributions Reveal Planning--Execution Dissociations}
Band-pair heatmaps (Figure \ref{fig:bandpair_heatmaps}) demonstrated that planning-phase contributions were broadly distributed across delta, theta, beta, and gamma interactions, reflecting engagement of large-scale coordination networks for object evaluation and motor preparation. In contrast, execution-phase interactions showed sharper and more condition-specific hotspots, particularly in $\beta$--$\gamma$, $\theta$--$\beta$, and $\beta$--$\alpha$ couplings. Such focal patterns align with sensorimotor demands of grasp stabilization, proprioceptive updating, and real-time object interaction. These dissociations illustrate that the brain transitions from distributed preparatory coupling to specialized high-frequency coordination once execution is initiated.

\subsection{Feature Selection and Dimensionality Reduction}
Permutation-derived feature subsets improved Planning-phase accuracies and preserved execution-phase performance relative to full-band models (Table \ref{tab:tableAcc}), indicating high redundancy within the full feature space. Only a small proportion of features carried the majority of discriminative information. This has important implications for real-time BCIs and neuroprosthetics, where compact, physiologically meaningful feature sets can reduce computation time and enhance generalization.

\subsection{Relation to Previous Work}
Prior bispectral studies have largely focused on motor imagery or limited spectral descriptors and have not examined executed grasping behavior \cite{sun2019advanced, saikia2011bispectrum, kotoky2014bispectrum, shahid2011bispectrum, das2016multiple, jin2020bispectrum, 11029483, 10782163, farhadi2026endtoendoptimizationbeliefpolicy}. To our knowledge, this is the first study to apply comprehensive cross-frequency bispectral analysis---including 25 band-pair interactions and both magnitude- and phase-based features---to natural reach-to-grasp execution. While direct comparisons are limited due to the novelty of this domain, our findings converge with invasive ECoG results showing enhanced high-frequency coherence during grasping, and with EEG studies linking beta--gamma modulations to fine motor control.

\subsection{Limitations and Future Directions}
Several limitations should be acknowledged. First, although the 16-channel EEG montage provided balanced cortical coverage, higher-density arrays may yield additional spatial resolution. Second, cross-bispectral connectivity between channels was not investigated; incorporating inter-channel nonlinear interactions may reveal broader network-level mechanisms. Third, the study involved only healthy adults, and future work should examine clinical populations such as stroke or spinal cord injury to assess translational relevance. Bispectral analysis offers physiologically interpretable insights into cross-frequency interactions (e.g., theta–beta coupling) while maintaining strong discriminative performance. Although the present study focused on classical machine-learning classifiers, integrating bispectral representations with deep learning models may further enhance performance while retaining interpretability.

% Finally, although classical machine-learning classifiers demonstrated strong discriminative power, integrating bispectral representations with deep learning models may further enhance performance and enable real-time decoding.

\section{Conclusion}
This study demonstrates that cross-frequency bispectral EEG analysis provides a powerful framework for characterizing nonlinear neural dynamics underlying executed reach-to-grasp behavior. By jointly examining magnitude- and phase-based bispectral features across multiple frequency band pairs, we show that execution execution is associated with stronger and more spectrally specific cross-frequency coupling than motor planning, particularly in $\beta$- and $\gamma$-driven interactions. The results further indicate that while planning engages a distributed set of nonlinear interactions, execution sharpens into more focal, task-relevant coupling patterns. These findings extend the application of bispectral analysis beyond motor imagery to ecologically valid grasping actions and highlight its potential for advancing neural decoding, brain–computer interfaces, and neurorehabilitation technologies.

\section{Acknowledgment}
\noindent The materials reported in this research project were supported by the NSF CAREER under award ID 2441496 and the NSF grant under award ID 2245558. Additionally, the project was supported by URI Foundation Grant on Medical Research and the Rhode Island INBRE program from the National Institute of General Medical Sciences of the NIH under grant number P20GM103430. 

\appendix
\section{}
\label{app1}
Additional figures, synthetic signal demonstrations, and validation analyses
are provided in the Supplementary Material. Code used in this study is
publicly available at:

\url{https://github.com/AbiriLab/Biosignal-Analysis-Bispectrum}.

%%       \cite{<label>} ==> [1]

%%
% Example citation, See \cite{lamport94}.





% %% For numbered reference style

%   \textit{\LaTeX: a document preparation system},
%   Addison Wesley, Massachusetts,

% \end{thebibliography}
\bibliographystyle{elsarticle-num}
\bibliography{main}

\end{document}

\begin{frontmatter}



\title{Supplementary Material for:\\ Cross-Frequency Bispectral EEG Analysis of Reach-to-Grasp Planning and Execution}

\author[label1]{Sima Ghafoori}
\author[label1]{Anna Cetera}
\author[label1]{Ali Rabiee}
\author[label1]{Mohammad Hasan Farhadi}
\author[label2]{Mariusz Furmanek}
\author[label1]{Yalda Shahriari}
\author[label1]{Reza Abiri}

\affiliation[label1]{
  organization={Department of Electrical, Computer and Biomedical Engineering, University of Rhode Island},
  city={Kingston},
  state={RI},
  country={USA}
}

\affiliation[label2]{
  organization={Department of Physical Therapy, University of Rhode Island},
  city={Kingston},
  state={RI},
  country={USA}
}

\end{frontmatter}



\section*{Supplementary Information}
This document provides additional methodological details and supporting analyses
that complement the main manuscript.

\section{Methods}
\label{sec2}
\subsection{Bispectral and Bicoherence Computation}

Bispectral analysis was used to characterize nonlinear cross-frequency
interactions by estimating the third-order spectrum of EEG signals.
For a signal $x(t)$, the bispectrum is defined as
\begin{equation}
    B(f_1,f_2)=\mathbb{E}\left[X(f_1)X(f_2)X^{*}(f_1+f_2)\right],
\end{equation}

where $X(f)$ denotes the Fourier transform and $\mathbb{E}[\cdot]$ represents
expectation across epochs. The bispectrum captures quadratic phase coupling
among frequency components and preserves both magnitude and phase information.

To enable comparison across frequencies and channels, the complex bicoherence
was computed as a normalized form of the bispectrum,
\begin{equation}
  b(f_1,f_2)=\frac{B(f_1,f_2)}
{\sqrt{\mathbb{E}[|X(f_1)X(f_2)|^2]\;\mathbb{E}[|X(f_1+f_2)|^2]}}.  
\end{equation}

This normalization bounds the magnitude of bicoherence between 0 and 1 while
retaining phase information, allowing assessment of both the strength and
directionality of nonlinear phase coupling. Bispectral and bicoherence estimates were obtained using a 256-point FFT with Hanning windows and 50\% overlap across time segments. Feature extraction and band-pair definitions are described in the main manuscript.

\subsubsection{Sum–Frequency and Phase–Locking Conditions}

The bispectrum at $(f_1,f_2)$ is defined according to equation (1), where $f_3 = f_1 + f_2$. From this definition two key rules follow \cite{nikias1993signal, sigl1994introduction, kovach2018bispectrum}:

\begin{enumerate}
  \item \textbf{Sum–frequency constraint}
    \begin{equation}
      f_3 = f_1 + f_2,\quad
      0 < f_1,\; f_2,\; f_3 \le \frac{f_s}{2}.
    \end{equation}
    If $f_1 + f_2$ exceeds the Nyquist limit ($f_s/2$), or if no energy is present at $f_3$, then $B(f_1,f_2) \approx 0$ \cite{nikias1993signal, sigl1994introduction}.

  \item \textbf{Phase–locking condition}
    \begin{equation}
      \angle B(f_1,f_2)
      = \angle X(f_1) + \angle X(f_2) - \angle X(f_3).
    \end{equation}
    A prominent bispectral peak at $(f_1, f_2)$ indicates that the phases satisfy
    \(\phi(f_1) + \phi(f_2) - \phi(f_1 + f_2) \approx 0\) \cite{sigl1994introduction, kovach2018bispectrum}.
\end{enumerate}

In particular, if two input tones at $f_1$ and $f_2$ do \emph{not} jointly generate or phase-lock with a component at $f_1 + f_2$, their (cross-)bispectrum vanishes \cite{nikias1993signal, sigl1994introduction}; that is,
\[
B(f_1, f_2) = 0 \quad \text{whenever the sum-frequency } f_1 + f_2 \text{ is absent or uncorrelated}.
\]

\subsection{Illustrative Examples with Synthetic Signals}
\begin{figure}[ht]
\centering
\includegraphics[width=\linewidth]{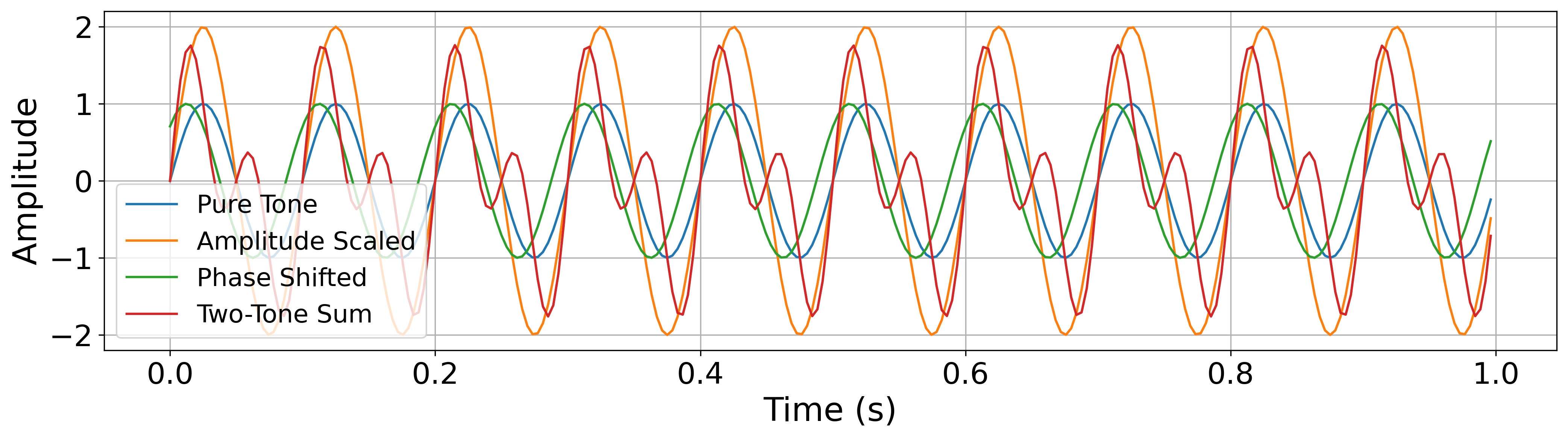}
\caption{Time-domain waveforms of four illustrative synthetic signals, each of 1\,s duration sampled at 256\,Hz.
\textbf{Pure Tone}: $f_0 = 10\,$Hz.
\textbf{Amplitude-Scaled}: $A = 2$, $f_0 = 10\,$Hz.
\textbf{Phase-Shifted}: $f_0 = 10\,$Hz, $\phi = \pi/4$.
\textbf{Two-Tone Sum}: $f_1 = 10\,$Hz, $f_2 = 20\,$Hz.}
\label{sin_signal}
\end{figure}

Before applying bispectral analysis to physiological data, we examine four simple signals. For each, we plot (i) the time-domain waveform (Figure \ref{sin_signal}), (ii) the amplitude bispectrum $|B(f_1,f_2)|$, and (iii) the phase bispectrum $\angle B(f_1,f_2)$.

\begin{figure}[ht]
  \centering
  \subfloat[][ \label{fig:suba}]{\includegraphics[width=0.49\linewidth]{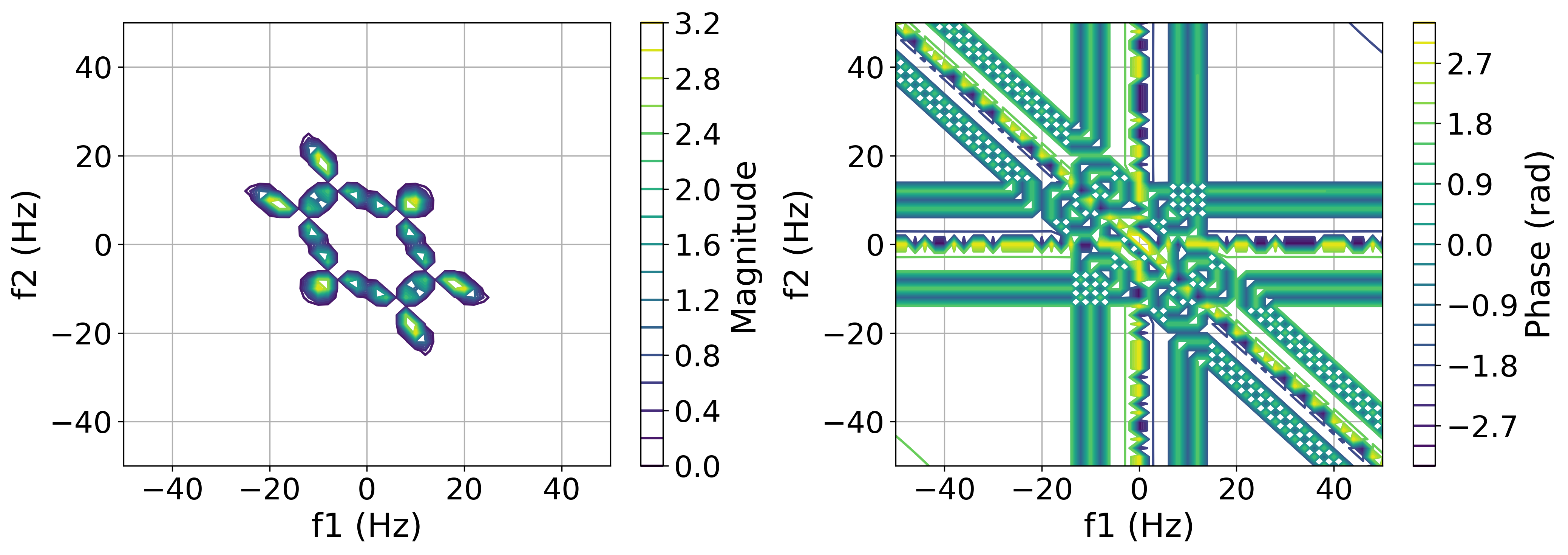}}
  % \hfill
  \subfloat[][ \label{fig:subb}]{\includegraphics[width=0.49\linewidth]{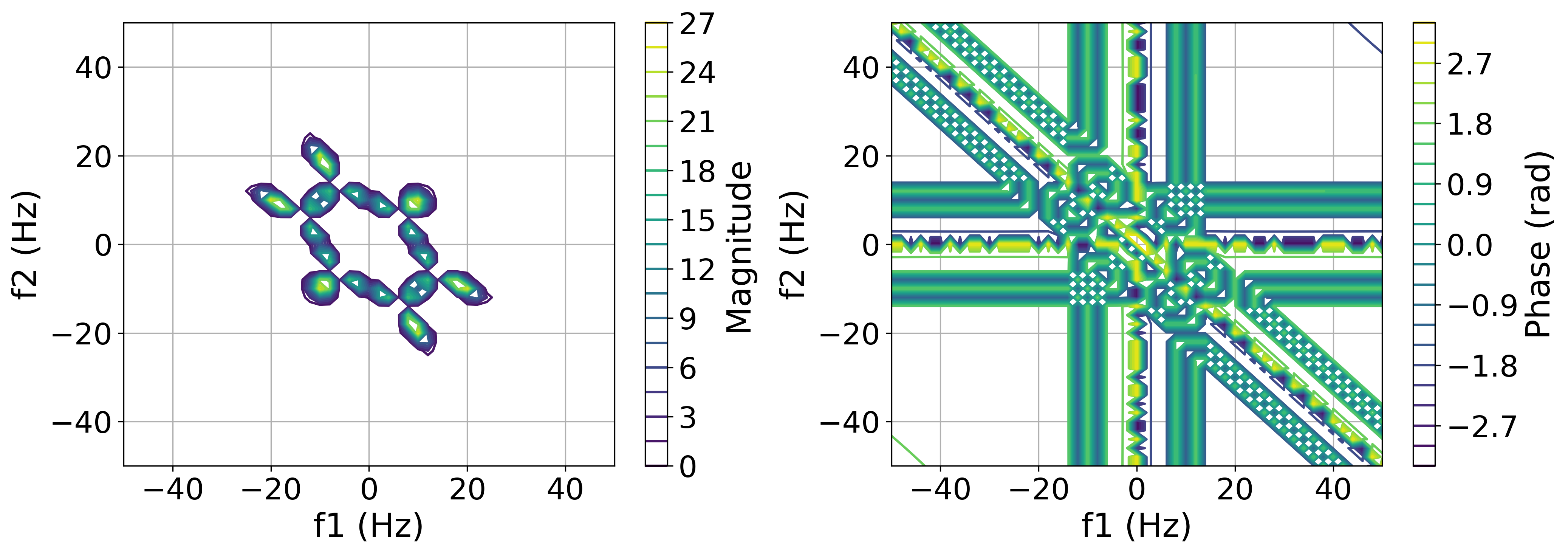}}\\
  \subfloat[][ \label{fig:subc}]{\includegraphics[width=0.49\linewidth]{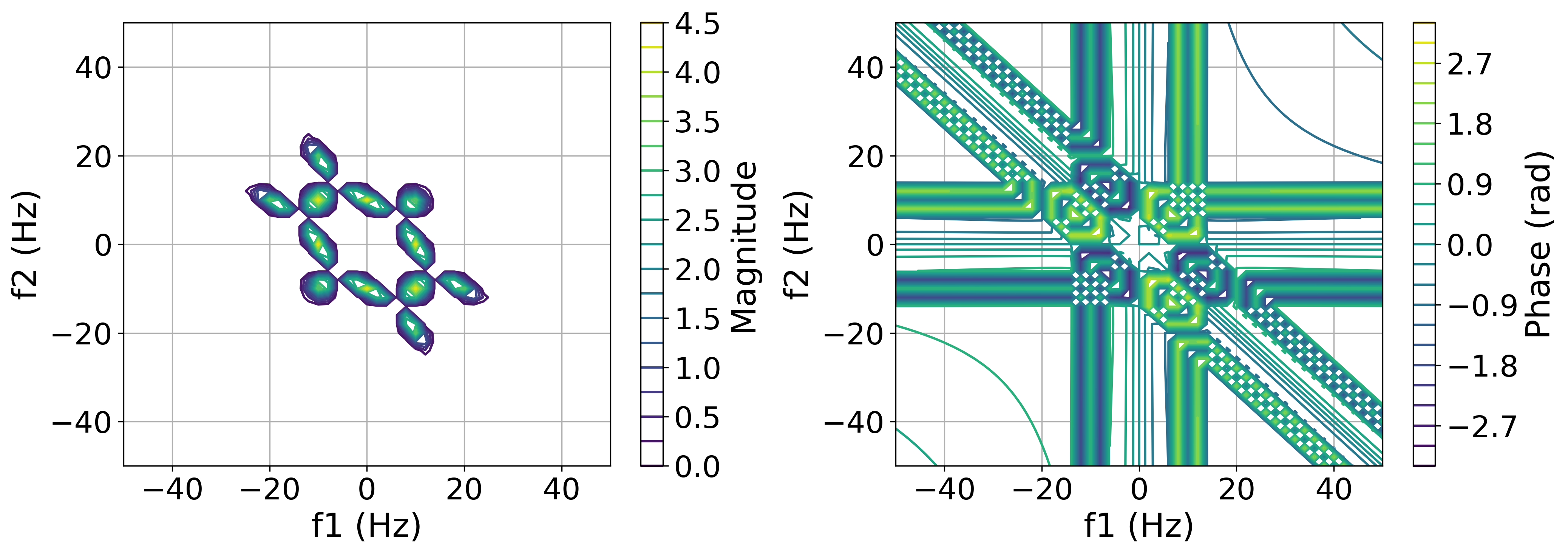}}
  % \hfill
  \subfloat[][ \label{fig:subd}]{\includegraphics[width=0.49\linewidth]{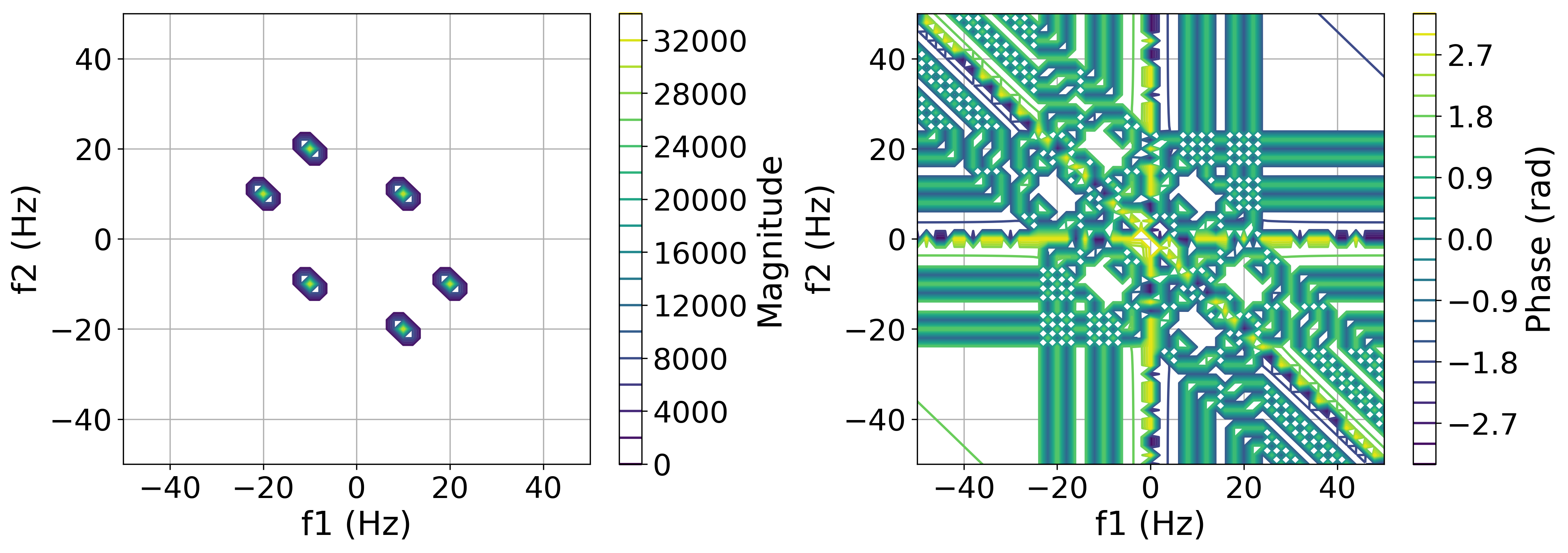}}
  \caption{Bispectral magnitude (left panel of each pair) and phase (right panel of each pair) for four synthetic signals: (a) Pure 10 Hz tone, (b) Amplitude-scaled tone, (c) Phase-shifted tone, and  (d) Two-tone sum (10 Hz + 20 Hz).}
  \label{sin_Bispect}
\end{figure}

\begin{itemize}
  \item \textbf{Pure Tone:} $x(t)=\sin(2\pi f_0 t)$.  
    Only one spectral line at $f_0$ and a constant phase at that bin. In the bispectral magnitude map (Figure \ref{sin_Bispect}-a, right), the pure 10 Hz sine exhibits a clear self–coupling peak of \textbf{3.1889} at $(f_1,f_2)=(10,10)$. Although an ideal linear tone should produce $B(10,10)=0$, our windowed FFT estimator and finite data length introduce leakage that manifests as this residual ridge at the sum frequency 20 Hz. The corresponding phase map (Figure \ref{sin_Bispect}-a, left) shows a phase of \textbf{–1.6652 rad} at that bin, reflecting the relative timing of the two 10 Hz inputs and the 20 Hz leakage term.  

    In fact, this global maximum magnitude of 3.1889 occurs at six symmetric locations: (10,10), (10,–20), (20,–10), (–20,10), (–10,20) and (–10,–10) Hz. At (10,10), (10,–20) and (–20,10) the phase is –1.6652 rad, while at (20,–10), (–10,20) and (–10,–10) it is +1.6652 rad, revealing the bispectrum’s conjugate symmetry.

  \item \textbf{Amplitude-Scaled Tone:} $x(t)=A\sin(2\pi f_0 t)$ with $A=2$.  
    The magnitude spectrum is scaled by $A$, while the phase spectrum remains unchanged. When the 10 Hz sine is doubled in amplitude, the bispectral magnitude peak at $(10,10)$ swells from 3.1889 to \textbf{25.5112} (Figure \ref{sin_Bispect}-b), an 8× increase exactly matching the $A^3$ scaling law (since $B\propto X(f_1)\,X(f_2)\,X^*(f_1+f_2)$). The phase map (Figure \ref{sin_Bispect}-b) again reports –1.6652 rad at $(10,10)$, unchanged from the pure-tone case.  

    That global maximum of 25.5112 likewise appears at the same six locations: (10,10), (10,–20), (20,–10), (–20,10), (–10,20), (–10,–10) Hz. Phases remain –1.6652 rad at the first three and +1.6652 rad at the latter three, confirming amplitude only amplifies magnitude.

  \item \textbf{Phase-Shifted Tone:} $x(t)=\sin(2\pi f_0 t + \phi)$ with $\phi=\pi/4$.  
    The magnitude spectrum matches the pure tone, but the phase at $f_0$ is offset by $-\phi$. Applying a global phase offset $\phi=\pi/4$ to the 10 Hz sine leaves its bispectral magnitude peak at \textbf{4.3667} but relocates the coupling to $(10,0)$ (Figure \ref{sin_Bispect}-c), i.e.\ the interaction between the 10 Hz line and the induced DC term from the offset. The phase at that peak (Figure \ref{sin_Bispect}-c) is \textbf{–0.0000 rad}, confirming that $\angle B(10,0)=\angle X(10)+\angle X(0)-\angle X(10)=0$.  

    Moreover, this maximum of 4.3667 occurs at six positions: (0,10), (0,–10), (10,0), (10,–10), (–10,0), (–10,10) Hz, and at each the phase is 0.0000 rad, illustrating that the pure phase shift introduces symmetric DC–frequency couplings without net phase change.

  \item \textbf{Two-Tone Sum:} $x(t)=\sin(2\pi f_1 t) + \sin(2\pi f_2 t)$ with $f_1=10\,$Hz, $f_2=20\,$Hz.  
    Two distinct spectral lines at $f_1$ and $f_2$; no quadratic coupling appears in the FFT magnitude. For the linear sum of 10 Hz and 20 Hz sinusoids, the bispectral magnitude map (Figure \ref{sin_Bispect}-d) again peaks at \textbf{32010.0103} at $(10,10)$, representing both the self-coupling of 10 Hz and its quadratic interaction that reinforces the 20 Hz component. The phase map (Figure \ref{sin_Bispect}-d) shows –1.5709 rad at $(10,10)$, indicating a quarter-cycle shift between the 10 Hz inputs and the resulting 20 Hz output.  

    That maximal 32010.0103 value is found at the same six symmetric points: (10,10), (10,–20), (20,–10), (–20,10), (–10,20), (–10,–10) Hz. Phases are –1.5709 rad at (10,10),(10,–20),(–20,10) and +1.5709 rad at (20,–10),(–10,20),(–10,–10), confirming the expected quadratic couplings at $f_1+f_2$.
\end{itemize}

\subsection{Detailed Feature Extraction and Formulation}
This subsection provides detailed definitions and mathematical formulations of the bispectral and bicoherence-based features used in this study.

\paragraph{Mean Bicoherence Magnitude.}
This feature captures the average strength of nonlinear coupling across the bicoherence matrix. It is computed as the mean of the absolute values of the complex bicoherence values:

\begin{equation}
    \mu_b = \frac{1}{N} \sum_{f_1} \sum_{f_2} \left| b(f_1, f_2) \right|
\end{equation}

where \( b(f_1, f_2) \) is the complex bicoherence at frequency pair \((f_1, f_2)\), and \(N\) is the total number of frequency pairs in the matrix. A higher \(\mu_b\) indicates more widespread and consistent nonlinear phase coupling across the frequency domain.

\paragraph{Maximum Bicoherence Magnitude.}
This feature represents the strongest observed coupling in the bicoherence matrix. It is defined as the maximum absolute value among all complex bicoherence entries:

\begin{equation}
    b_{\text{max}} = \max_{f_1, f_2} \left| b(f_1, f_2) \right|
\end{equation}

where \( b(f_1, f_2) \) denotes the complex bicoherence at frequency pair \((f_1, f_2)\). A higher \( b_{\text{max}} \) reflects the presence of a particularly dominant nonlinear interaction within the trial.

\paragraph{Sum of Bicoherence Magnitudes.}
This feature aggregates the total strength of nonlinear coupling by summing the absolute values of all elements in the bicoherence matrix:

\begin{equation}
    S_b = \sum_{f_1} \sum_{f_2} \left| b(f_1, f_2) \right|
\end{equation}

where \( b(f_1, f_2) \) is the complex bicoherence at frequency pair \((f_1, f_2)\). Unlike the mean, this summation reflects both the intensity and the spatial extent of nonlinear interactions, providing an energy-like measure of overall coupling across the matrix.

\paragraph{Entropy of Bicoherence Magnitudes.}
This feature quantifies the unpredictability or dispersion of nonlinear coupling strengths across the matrix. The bicoherence magnitudes are first normalized to form a probability distribution:

\begin{equation}
    p_{f_1, f_2} = \frac{|b(f_1, f_2)|}{\sum_{f_1} \sum_{f_2} |b(f_1, f_2)|}
\end{equation}

Shannon entropy is then computed as:

\begin{equation}
    H_b = -\sum_{f_1} \sum_{f_2} p_{f_1, f_2} \log(p_{f_1, f_2})
\end{equation}

A higher \( H_b \) indicates more uniform distribution of coupling strengths, whereas a lower value suggests that few frequency interactions dominate.

\paragraph{Mean Phase Angle.}
This feature captures the average phase relationship between interacting frequencies using circular statistics. Let \( \theta_{f_1, f_2} = \arg(b(f_1, f_2)) \) be the phase angle of each matrix entry. The mean phase is computed via:

\begin{equation}
    \bar{\theta} = \arg\left( \sum_{f_1} \sum_{f_2} e^{j \theta_{f_1, f_2}} \right)
\end{equation}

This quantity reflects the dominant direction of phase coupling across the matrix.

\paragraph{Phase Entropy.}
This feature measures the variability in the distribution of bicoherence phase angles. A histogram of phase values is constructed and normalized to form a probability distribution \( p_k \), where \(k\) indexes phase bins:

\begin{equation}
    H_\theta = -\sum_{k} p_k \log(p_k)
\end{equation}

Higher values of \( H_\theta \) indicate more desynchronized and distributed phase relationships, while lower values imply phase-locking around a dominant angle.

\paragraph{Circular Variance.}
This feature quantifies the dispersion of phase angles using circular variance, defined as:

\begin{equation}
    V_\theta = 1 - \left| \frac{1}{N} \sum_{f_1} \sum_{f_2} e^{j \theta_{f_1, f_2}} \right|
\end{equation}

where \(N\) is the number of matrix elements. A value close to 0 indicates tightly clustered phase values, while values near 1 reflect a broad spread.

\paragraph{Phase Concentration (R).}
Also known as the mean resultant length, this feature measures the concentration of phase angles around a preferred direction:

\begin{equation}
    R = \left| \frac{1}{N} \sum_{f_1} \sum_{f_2} e^{j \theta_{f_1, f_2}} \right|
\end{equation}

A higher \( R \) value suggests stronger phase-locking and more consistent phase coupling.
\begin{table}
    \centering
    \renewcommand{\arraystretch}{1.3}  % makes rows taller
    \begin{tabular}{ccccccc} 
       Driver Band   &  \( \delta \)&  \( \theta\)&  \( \alpha \)&  \( \beta \)&  \( \gamma \)& WB\\ \hline
 \multicolumn{7}{c}{Planning}\\ \hline
         Bottle vs. Empty&  75.00&  70.66&  70.00&  76.66&  76.33& 78.00\\
         Bottle vs. Pen&  69.66&  74.00&  70.00&  77.33&  76.00& 78.00\\
         Pen vs. Empty&  73.66&  73.33&  71.33&  75.33&  75.66& 78.00\\
         Multi&  58.44&  57.55&  55.78&  58.22&  61.55& 66.22\\ \hline
 \multicolumn{7}{c}{Movement}\\ \hline
 Bottle vs. Empty& 81.66& 82.00& 79.33& 91.00& 91.33&93.33\\
 Bottle vs. Pen& 74.33& 72.66& 70.00& 79.33& 75.00&76.33\\
 Pen vs. Empty& 86.66& 87.33& 86.00& 90.66& 92.00&94.66\\
 Multi& 65.77& 66.22& 62.66& 70.00& 72.22&78.66\\ \hline
    \end{tabular}
    \caption{Classification accuracies across frequency bands, whole-band features for both Planning and Movement phases. Values correspond to mean accuracies ($\pm$ standard deviation) across subjects maxima, for binary classification tasks as well as for the multiclass condition  planning and movement}
    \label{tableAcc}
\end{table}

\subsection{Supplementary Results}

\begin{figure*}[!t]
\centering
\includegraphics[width=0.98\textwidth]{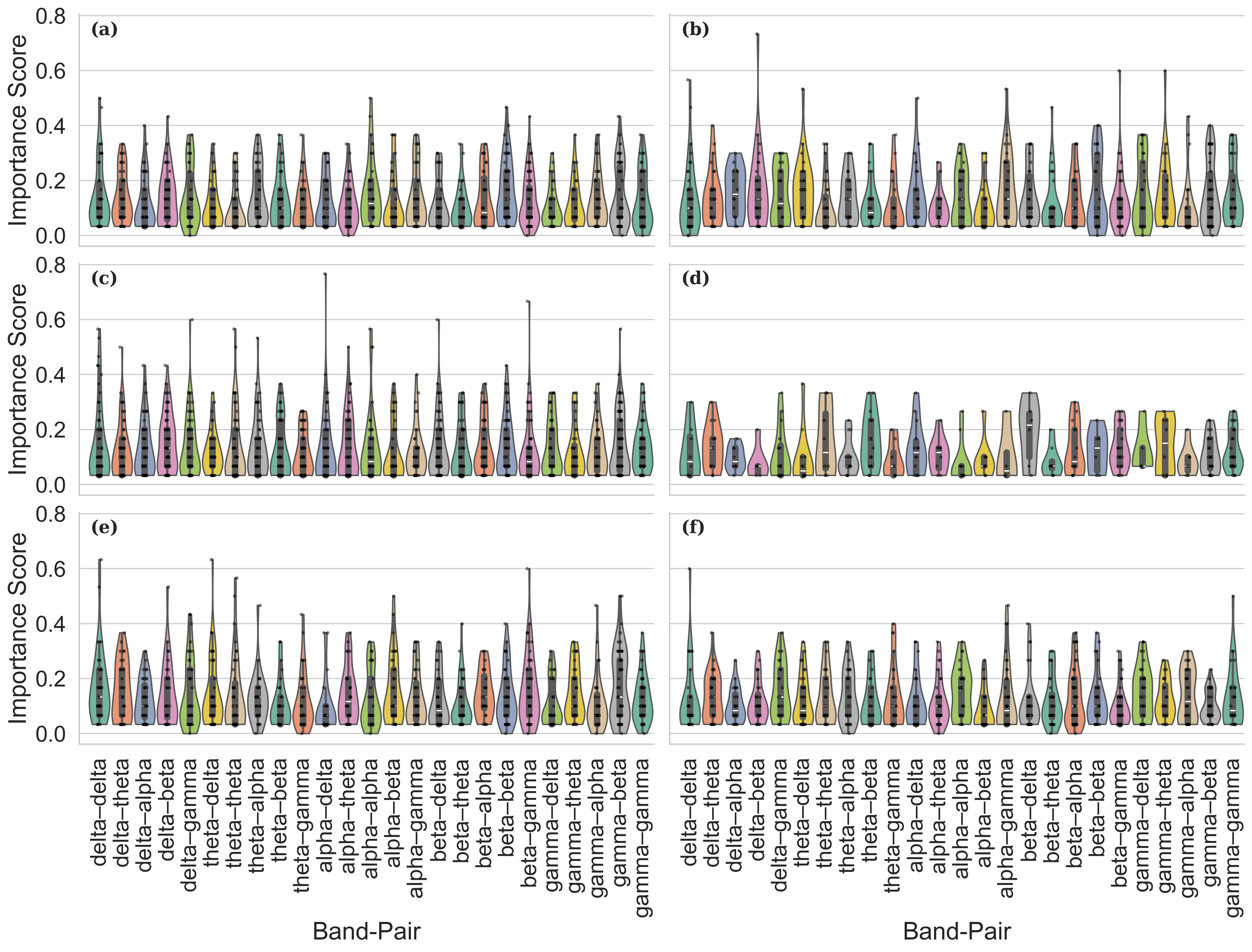}
\caption{Distribution of permutation-based feature importance scores across frequency band-pairs, tasks, and phases. Violin plots show cross-subject distributions of normalized importance values for the most informative bispectral EEG features. Rows correspond to classification tasks (Bottle vs.\ Empty, Pen vs.\ Empty, and Bottle vs.\ Pen), and columns correspond to the Planning and Movement phases. Each violin represents a specific frequency band-pair (x-axis), with the vertical axis indicating feature importance. Embedded boxplots denote the median and interquartile range.}
\label{selection_score}
\end{figure*}

Table \ref{tableAcc} reports classification accuracies across individual driver frequency
bands and whole-band (WB) features for both planning and movement phases. Reported values correspond to the mean of the maximum cross-validated training accuracy across subjects for each classification task. These values differ from those reported in the main manuscript, where performance is summarized using the mean of accuracies averaged across cross-validation folds and subjects.

Across all tasks, movement-phase accuracies are consistently higher than planning-phase accuracies. For example, in the Bottle vs.\ Empty task, planning accuracies range from 70.00--78.00\% across frequency bands, whereas movement accuracies increase to 79.33--93.33\%, with peak performance observed in the $\beta$ and $\gamma$ bands. Similar trends are observed for Bottle vs.\ Pen and Pen vs.\ Empty tasks, where movement-phase accuracies reach up to
94.66\% in high-frequency bands.

This table is included to provide a complete band-wise numerical reference and to report the upper-bound unseen performance achieved within each frequency band, complementing the averaged unseen performance results presented in the main manuscript.

Figure \ref{selection_score} summarizes the distribution of permutation-based feature importance
scores across frequency band-pairs for each classification task and phase. During the planning phase, importance values are broadly distributed across band-pairs, with relatively low medians and substantial overlap between distributions, indicating diffuse contributions from multiple cross- frequency interactions. In contrast, the movement phase shows higher median importance
scores and longer upper tails for several band-pairs, reflecting stronger and more consistent contributions during execution.

Across tasks, band-pairs involving higher frequencies, particularly those including $\beta$ and $\gamma$ bands, exhibit elevated importance during the movement phase, whereas planning-phase importance remains more evenly spread across low- and high-frequency combinations. Multi-class discrimination shows lower overall importance values and greater overlap across band-pairs compared to binary tasks, consistent with increased task complexity. Together, these distributions provide a detailed view of how nonlinear cross-frequency interactions contribute to decoding performance across tasks and movement phases.

\bibliographystyle{elsarticle-num}
\bibliography{main}